\documentclass[aps,prd,nofootinbib,amsfonts,amssymb,amsmath,letterpaper,showkeys,10pt,reprint,floatfix]{revtex4-2}

\usepackage{tensor,mathtools,graphicx,enumitem} 
\usepackage[colorlinks=true,allcolors=blue]{hyperref} 

\usepackage[brazil,english]{babel}
\usepackage{float}
\usepackage{amssymb,amsmath,amsthm,enumitem}
\usepackage[scr=rsfs]{mathalpha}
\usepackage{empheq}
\usepackage{graphicx}
\usepackage{relsize}
\usepackage{cancel}
\usepackage{verbatim} 
\usepackage{bbold}
\usepackage{braket}
\usepackage{xcolor}
\usepackage{tabularx}
\usepackage{booktabs}
\usepackage{indentfirst}
\usepackage{hyperref}
\usepackage{wrapfig}
\usepackage[font={small, stretch=1.0} , justification=justified]{caption}

\pdfoutput=1 

\newcommand{\Tr}{\operatorname{Tr}}   
\newcommand{\dket}[1]{\ket{#1}\!\rangle}   
\newcommand{\Pket}{\ket{\Psi}\!\rangle}    
\newcommand{\dbra}[1]{\langle\!\bra{#1}}   
\newcommand{\ketbra}[2]{\ket{#1}\!\bra{#2}}

\begin{document}

\title{A spacetime-covariant approach to inertial and accelerated quantum clocks in first-quantization}



\author{Eduardo A. B. Oliveira}
\email{eduardo.amancio@unesp.br}
\affiliation{Instituto de F\'isica Te\'orica, Universidade Estadual Paulista, \\ Rua Dr. Bento Teobaldo Ferraz, 271 - Bloco II, 01140-070 S\~ao Paulo, S\~ao Paulo, Brazil}

\author{Andr\'e G. S. Landulfo}
\email{andre.landulfo@ufabc.edu.br} 
\affiliation{Centro de Ci\^encias Naturais e Humanas, Universidade Federal do
ABC,\\  Avenida dos Estados, 5001, 09210-580 Santo André, S\~ao Paulo, Brazil}


\begin{abstract}
It is expected that a quantum theory of gravity will radically alter our current notion of  spacetime geometry. However, contrary to what was commonly assumed for many decades, quantum gravity effects could manifest in scales much larger than the Planck scale, provided that there is enough coherence in the superposition of geometries. Quantum Clocks, i.e., quantum mechanical systems whose internal dynamics can keep track of proper-time lapses, are a very promising tool for probing such low-energy quantum gravity effects. In this work, we contribute to this subject by proposing a spacetime-covariant formalism to describe clocks in first quantization. In particular, we account for the possibility of dynamically accelerated clocks via suitable couplings with external fields. We find that a particular decomposition of the (quadratic) clock Hamiltonian into positive- and negative-mass sectors, when attainable, enables one to compute the evolution of the system directly in terms of the clock's proper-time while maintaining explicit covariance. When this decomposition is possible, the evolution obtained is always unitary, even with couplings with external fields used to, e.g., accelerate the clocks. We then apply this formulation to compute the joint time evolution of a pair of quantum clocks in two cases: (i) inertial clocks with relative motion and (ii) charged clocks accelerated by a uniform magnetic field.  In both cases, when our clocks are prepared in coherent states, we find that the density matrix $\rho_{\tau_2|\tau_1}$ describing time measurements of the two clocks yields not only conditional probabilities whose peaks match exactly the classical expected values for time dilation, but also yields coherent quantum fluctuations around that peak, with a profile which is either (i) a pure Gaussian or (ii) a Gaussian combined with a periodic modulation (ii).
\end{abstract}


\maketitle

\section{Introduction} \label{sec:introduction} 

 Developing a satisfactory theory of quantum gravity is arguably one of the greatest challenges of contemporary physics. Even though many top-down approaches have presented us with candidate theories (such as Strings or Loop Quantum Gravity), none have so far presented compelling experimental predictions to favor their adoption. One obvious reason is that typical quantum-gravitational phenomena are expected to become relevant only near the Planck scale, i.e., distances smaller than the Planck length $l_p\sim 10^{-43}$ {\rm light-sec} or, equivalently, energies larger than Planck energy $E_p\sim 10^{19}$ {\rm GeV}, 14 orders of magnitude above the highest energies we can currently probe on any earth-based experiments. 
 
 However, in recent years, it has become increasingly clear that quantum-gravitational phenomena could be relevant in another complementary regime, namely, when there are large coherent (relative) fluctuations of the stress tensor: $\braket{T_{\mu\nu}^2} \gg \braket{T_{\mu\nu}}^2 $. This could be achieved in situations as `ordinary' as a spatial superposition of a non-relativistic massive particle. Indeed, a plethora of work both in gravity/quantum field theory in curved spacetimes and quantum foundations contexts (see, e.g., Refs.~\cite{waldbruck, waldbruck2, wald, BRovelli, ofek}) have convincingly argued that such regimes cannot be consistently described by quantum matter in a classical spacetime, such as semiclassical gravity\footnote{However, some authors argue that consistency is still possible for statistical theories of gravity with statistical fluctuations. See \textit{e..g} \cite{oppenheim}.}. Hence, this would yield genuinely quantum-gravitational effects measurable at arbitrarily low energies. 
 
 Currently, there is a lot of experimental effort to design and implement experiments that are sensitive enough to measure such effects. They are mainly focused on the gravitationally induced entanglement between two massive particles in quantum superpositions \cite{MV, Bose}. Although there are many challenges involved in preparing sufficiently massive particles in coherent superpositions for long enough to entangle them gravitationally, this is expected to be feasible in the next few decades \cite{aspelmeyeressay, aspelmeyertalk}.

On the theoretical side, a lot of effort has been put into understanding the low-energy regime of nonclassical spacetime geometry, especially how to describe it operationally in terms of physical probes and their corresponding observables \cite{bruckAV, bruckTBell, rovelliP1, rovelliP2, QFT-QRFs, rick, jan, foo1, foo2, foo3}. Classically, this can be done by means of (ideal) measurements on ideal clocks~\cite{geroch, FeliceBini, waldGR}, i.e., pointlike massive particles following timelike worldlines with some internal degree of freedom (DoF). By measuring the value of such a DoF one is able to `read off´ the proper time elapsed along any worldline interval. One can reconstruct the geometry of spacetime on any region by combining sufficiently many clock readings and, thereby, test the predictions of General Relativity or any alternative classical gravity theory. For non-classical spacetimes, quantum clocks -- i.e., quantum mechanical systems with internal DoFs that allow one to measure proper-time lapses -- are quite natural candidates for playing the role of physical probes whose measurements allow us to reconstruct the spacetime geometry. This would enable one to explore the interplay between quantum phenomena and gravity and test predictions of quantized theories of gravity. 

Classical relativistic effects in quantum clocks are well known: both kinematic and gravitational time dilation lead to decoherence of center-of-mass (CM) DoFs for clocks in superpositions of different spatial trajectories~\cite{bruckD1, bruckD2, unruh}. More interestingly, quantum clocks have proven to be a valuable conceptual and quantitative resource to explore phenomena either in the complete absence of a classical background spacetime or in situations where one tries to wash out any explicit reference to a prior spacetime structure and build a description directly in terms of physics observables measurable by physical probes. Indeed, in many cases, one recovers emergent notions of spacetime geometry or time evolution from these observables~\cite {PW, giovanetti, flaminiaSTQRFs}.

Much of the inspiration for such relational approaches comes directly from General Relativity (GR) and Canonical Quantum Gravity (CQG). As these theories lack any preferred background structure, they have several gauge symmetries associated with diffeomorphism invariance. This leads to many kinematic constraints for the (physical) phase space of the classical theory and, thus, for the corresponding (physical) Hilbert space of the quantized theory. A paradigmatic example is the Hamiltonian constraint, $H \approx 0$ for GR and $H\ket{\Psi}\rangle=0$ for CQG (the latter is known as the Wheeler-DeWitt equation~\cite{CQG}), expressing the fact that (global) time evolution can be seen as a mere gauge transformation. This leads to the so-called ``problem of time'', referring to the seeming contradiction that systems obeying this constraint should not evolve at all while we obviously observe nontrivial time evolution for all kinds of systems. The solution to this ``paradox'' comes from the fact that, while the (global) time evolution of all systems is equivalent to a mere diffeomorphism, there is a physically meaningful (gauge-invariant) way to define time evolution relationally through correlations between two or more physical systems. To put it briefly, systems evolve with respect to one another.


Inspired by such issues, the framework of constrained systems with gauge freedom has been widely used in the field of quantum foundations and its interface with gravity~\cite{PN1, PN2, bruckQRF, bruckQCG, flaminiaSTQRFs, PW, giovanetti, bruckQCG, flaminiaSTQRFs, smithQC1, trinity, trinity2, aharonov, evan, flafis, kasia}. This is often done in the form of toy models with a finite number of DoFs (for an interesting conceptual discussion of the role of gauge symmetries in modern physical theories, see~\cite{rovGauge}). In these models, the gauge freedom is introduced to enforce an underlying symmetry, making the theory completely relational (background independent). As a result, the only gauge-invariant observables will be relational ones. In particular, for models with a Hamiltonian constraint, $H\Pket=0$, one does not have any background notion of time evolution, but one can construct an emergent notion of time evolution relationally via the so-called Page-Wooters (PW) mechanism~\cite{PW, giovanetti, bruckQCG, flaminiaSTQRFs, smithQC1, IQC}, 
This is done by splitting the total system into a `clock' plus `remaining' DoFs and then evaluating the `remaining' DoFs conditioned on the clock's state. With such a relation, the constraint equation for the global system  becomes a dynamical equation for the remaining subsystem, describing how its conditional states change with respect to the clock's time.

Most of the toy models employing the PW framework are nonrelativistic (sometimes with the insertion of relativistic corrections). An important example of this approach is given in Ref.~\cite{aharonov}, where the authors claim that accelerated quantum clocks generally describe a nonunitary time evolution for their CM DoFs. This sometimes makes the Hamiltonian constraints that they introduce somewhat ad hoc and failing to consider important relativistic features\footnote{Relativistic corrections inserted by hand might differ from the corrections appearing in the weakly relativistic limit of a system covariantly constrained.}. In relativistic particle mechanics, on the other hand, spacetime-covariant Hamiltonian constraints appear naturally. These systems have internal symmetries, namely reparametrization invariance in each particle's worldlines, that play an analogous role to diffeomorphism invariance in GR. There are already a few works modeling relativistic quantum clocks using these relativistic constraints arising from reparametrization symmetry~\cite{flaminiaSTQRFs, smithQC1, trinity2}\footnote{In \cite{nahuel1,nahuel2} the authors also consider relativistic history states arising from covariant constraints. {However (except for a brief mention in \cite{nahuel1}) they do not consider an internal clock subspace to define temporal observables or time evolution; rather, they give a covariant constrained treatment to the external time variable, recovering an external time evolution for states in subspaces of well defined masses.}}. By quantizing these covariantly constrained theories appropriately, one can model relativistic effects from first principles and take the weakly relativistic limit to compare its results with those obtained by inserting relativistic corrections into nonrelativistic models.

There are still important limitations and issues to address regarding relativistic quantum clocks. One quite fundamental issue is that most existing works use ideal quantum clocks. These are clocks whose internal Hamiltonian $H_C$ has a continuous unbounded spectrum, $\sigma(H_C)=\mathbb{R}$, allowing one to define a self-adjoint time operator $T_C$ conjugated to it, i.e., $[T_C,H_C]=i$. The  eigenstates $\ket{t},\,t\!\in\!\mathbb{R}$, of $T_C$ have well defined proper times. Aside from the usual issues of having a Hamiltonian that is unbounded from below (which was the main reason for Pauli's famous criticism to a time operator in quantum mechanics~\cite{pauli}, later refined by Unruh and Wald \cite{UnruhWald}), such as runaway solutions, the sharply localized time states $\ket{t}$ require an arbitrarily large spread of the clocks' internal energy, in accordance with the time-energy uncertainty relation $\Delta t \geq 1/(2\Delta E)$.  While this is fine in nonrelativistic mechanics it leads to fundamental issues both in special-relativistic and gravitational scenarios. In the former, the issues are similar to those of relativistic particle mechanics, where one cannot define arbitrarily localized states for a particle, as there is a threshold when momentum uncertainty becomes comparable with the particle's mass, leading to particle creation. Furthermore, when gravity comes into play, an arbitrarily localized clock with increasingly large energies produces increasingly large distortions in the very spacetime whose intervals it is trying to measure. In the extreme case, the clock's energy would collapse the spacetime around it in a black hole. We can circumvent such problems by employing the so-called temporal POVMs~\cite{trinity, trinity2}. This models physical clocks with finite precision and obtains the expected bound for physical quantum clocks: they cannot be simultaneously localized enough and precise enough to be able to measure spacetime intervals more precise than Planck scales, $t_P\sim 10^{-43} {\rm s}.$ 

In the present work, we give our contribution to this emerging field by proposing a new fully covariant framework for first-quantized relativistic quantum clocks, which attempts to tackle many of the aforementioned issues and limitations. By considering the decomposition of the (quadratic) Hamiltonian constraint in positive and negative \textit{mass} sectors, rather than energy sectors, we are able to define time evolution directly with respect to the clocks' internal proper time while maintaining explicit covariance. Furthermore, by considering physical clocks with a bounded internal spectrum $\sigma(H_C) \subset [E_{min},E_{max}]$, we can make sure all physical states of the theory lie in the positive mass sector, such that the quadratic and linear constraints span equivalent physical Hilbert Spaces. Finally, the framework can be easily extended to consider dynamically accelerated clocks, which we do by introducing covariant couplings with external fields, such as a scalar field and the electromagnetic field. This, in particular, will enable us to address the claim in~\cite{aharonov} that accelerated clocks should yield nonunitary time evolution, as our model predicts unitary evolution for any center-of-mass coupling that does not directly affect internal energies of the clock (such as an electromagnetic coupling). 

We find this framework particularly appealing to a relativistic-oriented community, and we hope it helps to bridge approaches in the interface between gravity and quantum foundations. We also think it can provide a complementary angle to many features appearing in the existing literature for (relativistic) quantum clocks and shed some new light on their results. {In particular, we revisit the notion of ``quantum time dilation'' (QTD) introduced in \cite{smithQC1}, a feature that quantifies how quantum coherence may affect proper time observables and the statistics of their measurements. While their formulation defines QTD through (coherence-induced) shifts in conditional probabilities in proper time measurements for 2 clocks $P(T_2\!=\tau_2|T_1\!=\!\tau_1)$, here, we propose a different definition based on a conditional density operator $\rho_{\tau_2|\tau_1}$, which quantifies the coherence in the superpositions of proper time itself. We find the term ``quantum time dilation'' well-suited for this definition, as it is precisely this coherence that allows, for instance, the physical realization of scenarios with indefinite causal order \cite{bruckTBell, ICO1, ICO2}.}

The outline of this paper is as follows. In Sec.~ \ref{sect:CovPM} we write down the covariant Lagrangian description of our system of interest: a system of massive particles with internal DoF serving as clocks (possibly coupled with external fields). We emphasize the constraint that such systems inherit in the Hamiltonian formalism which can be decomposed in positive- and negative-mass sectors. 
In Sec. \ref{sec:Quantization}, we discuss a method for covariantly quantizing this system and derive the properties of the internal time evolution conditioned by the internal states of a chosen fiducial reference clock. We also consider the case of  multiple clocks, where we have to impose extra constraints on the physical states of our theory to obtain a commensurable notion of time evolution. In Sec.~\ref{sect:Applications}, we apply our formalism to a few simple systems, namely, {\bf (1)} a pair of free quantum clocks with some relative velocity and {\bf (2)} a pair of charged quantum clocks in an uniform magnetic field, for which we explicitly compute their time evolution and the conditional density operators $\rho_{\tau_2|\tau_1}$, from which we find coherent time dilation. In Sec.~\ref{sec:conclusions} we give our final remarks.

We work in Minkowski spacetime with metric signature $(+,-,-,-)$ and use natural units where $\hbar = c = 1$. 

\section{A spacetime-covariant treatment of relativistic particle mechanics}
\label{sect:CovPM}

In order to propose our covariant formalism to describe the evolution of relativistic quantum systems with respect to quantum clocks, it will be useful to first describe classical covariant framework. Although there are some works in the literature that give a spacetime-covariant (classical) treatment for relativistic particle mechanics \cite{cromer, marnelius, rovelliQMWT, rovelliCQM}, this is not a widespread subject. For this reason, and also because many of the features of our quantum clocks arise very naturally with a covariant treatment, we shall give the subject a more detailed description here.

\subsection{Symmetries and constraints} \label{symmetries}

Let us consider some (fixed) inertial coordinate system  $\{x^\mu\}, \;{\mu=0,1,2,3}$, in which the CM DoF of a relativistic particle follows a worldline $x^\mu(\tau)$, with $\tau$ being its proper time. The action that describes the dynamics of the CM DoF, parametrized by the proper time $\tau$, is then

\begin{align}
    S = \int d\tau \mathcal{L}(x^\mu, u^\mu),
\end{align}
where $\mathcal{L}$ is the Lagrangian and $u^\mu \equiv {dx^\mu}/{d\tau}$ is the normalized 4-velocity. We note, however, that treating the 4 components of $x^\mu$ (and $u^\mu$) on an equal footing, as if they were completely independent variables, introduces extra spurious DoFs (now we have 4 DoFs per particle rather than 3). In particular, when we derive Euler-Lagrange equations, we obtain

\begin{align}
 \delta S = 0 \qquad \Rightarrow \qquad \frac{\partial \mathcal{L}}{\partial x^\mu} - \frac{d}{d\tau} \biggl( \frac{\partial \mathcal{L}}{\partial u^\mu} \biggl) = 0 \label{cov EL}
\end{align}
but must also impose the normalization condition

\begin{align}
    u_\mu u^\mu = 1 \qquad \Rightarrow \qquad a_\mu u^\mu =0 \label{4-normalization},
\end{align}
where $a^\mu \equiv {du^\mu}/{d\tau}$ is the 4-acceleration, as for  an arbitrary choice of $\mathcal{L}$, Eq.~ \eqref{cov EL} generally yields $a_\mu u^\mu \neq 0$. 

Geometrically, the constraint in Eq. \eqref{4-normalization} expresses the condition that $\tau$ is an affine parameter in the particle's worldline. Interestingly, with an appropriate choice of action, one can incorporate this geometrical constraint covariantly as a gauge fixing. This is done as follows. Instead of working \textit{a priori} with the affine parameter $\tau$, we write the action on an arbitrary parametrization $s(\tau)$ as


\begin{align}
    S = \int ds L(x^\mu, U^\mu),
\end{align}
where $L$ and $U^\mu \equiv {dx^\mu}/{ds}$ are the Lagrangian and generalized 4-velocity with respect to the arbitrary parameter $s$, respectively. The norm of $U^\mu$ will then be given by
\begin{align}
    \sqrt{U_\mu U^\mu} = \displaystyle{\frac{d\tau}{ds}} \equiv  \gamma(s).
\end{align}

For the action $S$ to be a scalar (invariant) under reparametrization choice, we must have that $L$, which we will call the \textit{covariant Lagrangian}, is a {scalar density}, transforming covariantly under reparametrizations $s \rightarrow s'(s)$, with $s'(s)$ being any smooth monotonic function, as

\begin{align}
    ds \,L(x^\mu,U^\mu) &= ds' \,L'(x^\mu,{U'}^\mu) 
\end{align}
and thus 
\begin{align} \label{transfL}
    ds' \biggl[\frac{ds}{ds'} L(x^\mu,U^\mu) \biggl] &= ds' \biggl[ L' \biggl(x^\mu, \frac{ds}{ds'}U^\mu \biggl) \biggl].
\end{align}
As Eq.~(\ref{transfL}) must hold for any reparametrization  we conclude that $L$ must be an homogeneous function of degree 1 in $U^\mu,$ i.e., it satisfies

\begin{align}
    L(x^\mu, \lambda U^\mu) = \lambda L(x^\mu, U^\mu) \label{Lagrangian-covariance}
\end{align}
for any $\lambda\in \mathbb{R}_+$. This very specific dependence reduces dramatically possible choices of a relativistic Lagrangian, and automatically exclude those that violate Eq.~\eqref{4-normalization}. In particular, for a free particle, the action that satisfies these criteria and recovers the usual (geodesic) equations  of motion for a relativistic particle is

\begin{align}
    S_{\rm free} = \int d\tau \, (-m \sqrt{ u_\mu u^\mu }) = \int ds (-m \sqrt{ U_\mu U^\mu }). \label{S_FP(good)}
\end{align}

Formulated this way, it becomes manifest that the extra degree of freedom arising in the covariant formulation turns out to be a gauge variable associated with reparametrization freedom. Interestingly, this freedom is associated with the physical arbitrariness of a time variable which is extrinsic to the dynamics (in the present case, just an arbitrary length coordinate in the particle's worldline). The specification of an operational time variable is ultimately bound to the dynamics, with a physical system serving as a clock. We will explore this in the next subsection.

Having presented a covariant Lagrangian formulation, we can derive the corresponding covariant Hamiltonian formulation in the usual fashion. We define the conjugate momentum $\mathbb{P}_\mu$ in any given parametrization $s$ as

\begin{align}
   \mathbb{P}_\mu \equiv \frac{\partial L}{ \partial U^\mu } = \frac{\partial \mathcal{L}}{ \partial u^\mu }.
\end{align}
We note that, due to Eq. \eqref{Lagrangian-covariance}, $\mathbb{P}_\mu$ is parametrization-invariant. It is easy to verify that there is a primary constraint in our covariant phase space $\mathcal{P}_{phys}$, as the 4-components of $\mathbb{P}^\mu$ will not be independent. By using Eq.~(\ref{Lagrangian-covariance}), we immediately have (by Euler´s theorem) that its canonically conjugated Hamiltonian is constrained to vanish 

\begin{align}
    H \equiv \mathbb{P}_\mu U^\mu - L \approx 0,
\end{align}
where the symbol $\approx$ denotes a weak equality, i.e., an equality that holds in the constraint surface $\mathcal{P}_{phys}\subset \mathcal{P}_{kin}$, also known as the physical phase space, containing all possible values of $(x^\mu,\mathbb{P}_\mu)$ derived from our Lagrangian\footnote{For more details on constrained systems,  see~\cite{teitelboim, trinity, QCS2, QCS3, QCS4, QCS5}.}. To illustrate this, let us consider the free particle described by the action~(\ref{S_FP(good)}). The conjugate momentum is then be given by 
\begin{equation}
     \mathbb{P}_\mu = \frac{-m U_\mu}{\sqrt{U_\alpha U^\alpha}} \equiv -P_\mu,
\end{equation}
where $P_\mu$ is the (mechanical) 4-momemtum of the free particle, and the Hamiltonian constraint is just

\begin{align}
    H = \gamma(s) \Bigl(- \frac{P_\mu P^\mu}{m} + m \Bigl) \approx 0.
\end{align} 
By imposing the above constraint, one recovers the familiar relativistic dispersion relation
\begin{align}
    \qquad P_\mu P^\mu \approx m^2.
\end{align}

Finally, for $N$ noninteracting particles, the total action is just the sum of the action of each particle

\begin{align}
    S = \sum_{I=1}^N S_I = \sum_{I=1}^N \int ds_IL_I(x^\mu_I,U^\mu_I),
\end{align}
with $x^\mu_I$ and $U^\mu_I$ being the worldline and generalized 4-velocity of each particle. We note that each  $S_I$ has an independent reparametrization freedom along its worldline. In turn, for the Hamiltonian formulation, we correspondingly obtain $N$ independent constraints 
\begin{align}
    H_I \equiv  \mathbb{P}^I_\mu U_I^\mu - L_I  \approx 0 \label{independent local constraints}
\end{align}
(no sum on $I$) in covariant phase space.

\subsection{Incorporating internal DoFs} \label{iDOFs}

So far, our dynamical description encompassed only the CM DoF. However, to serve as physical clocks, our particles must also be endowed with some internal structure, whose configurations allow us to infer the elapsed proper-time intervals throughout their  worldlines. Such internal structure could be either a truly extensionless DoF (such as the spin of a particle), or something whose spatial extension is negligible in the scales considered. 
 In both cases, we shall only be concerned with positions as codified in the CM variables and treat this internal structure through an (extensionless) internal variable $q$.

In our covariant formulation, it is straightforward to incorporate this internal DoF properly in an extended action

\begin{align}
    S = \int d\tau \, \mathcal{L}(x^\mu, u^\mu; q, q')= \int ds \, L(x^\mu, U^\mu; q, \dot{q}),
\end{align}
where we defined the generalized internal velocities $q' \equiv {dq}/{d\tau}$ and $\dot{q}\equiv{dq}/{ds}$.

In the simplest case, the Lagrangian $\mathcal{L}$ is separable in internal and external terms as
\begin{equation}
\mathcal{L} = \mathcal{L}_{ext}(x^\mu, u^\mu) + \mathcal{L}_{int}(q, q'),
\end{equation}
such that $q$ will depend on the CM dynamics only through the proper time. (This is, of course, not the only logical possibility: one could have, for example, that $q$ is a magnetic momentum and the particle could be immersed in an inhomogeneous magnetic field $\mathbf{B}(x)$, which would yield $\mathcal{L}_{int}=\mathcal{L}_{int}(x,q)$). This separability is precisely what makes $q$ useful as a clock, since the internal evolution is given simply by

\begin{align}
    \frac{d}{d\tau} \biggl( \frac{\partial \mathcal{L}_{int}}{\partial q'} \biggl) - \frac{\partial \mathcal{L}_{int}}{\partial q} = 0.
\end{align}
From this, we can derive the internal state $q(\tau)$ along the particle's worldline. For a monotonic internal DoF, we can invert this relation to recover $\tau$ as a function of $q$, $\tau(q)$. More generally, such inversion can only be done locally as, for instance, in the paradigmatic examples where $q$ describes a harmonic oscillator or a pendulum (whose trajectories are periodic). For quantized systems, one also recovers periodic internal dynamics when the internal Hamiltonian has a regular discrete spectrum. For a thorough review on the possibilities for the phase space of the clock and corresponding implications for the time variables, see \cite{trinity}.


Finally, let us settle our notation and define a few quantities in the internal dynamics. Just as with external variables, we have the covariant internal Lagrangian

\begin{align}
    L_{int}(q,\dot{q};U^\mu) \equiv \sqrt{U_\mu U^\mu} \,\mathcal{L}_{int}(q,q')
\end{align}
Then, the internal conjugate momentum $P_q$ will be given by

\begin{align}
    P_q \equiv \frac{\partial {L}_{int}}{\partial \dot{q}} = \frac{\partial \mathcal{L}_{int}}{\partial q'},
\end{align}
which, just like $\mathbb{P}_\mu$, is invariant under reparametrizations. By using $P_q$, we can build both the proper and covariant internal Hamiltonians, $\mathcal{H}_q$ and $H_q$ as

\begin{align}
    \mathcal{H}_q \equiv P_q q' - \mathcal{L}_{int},\label{Hqa}
\end{align}
and 
\begin{align}
\qquad H_q \equiv P_q \dot{q} - L_{int} = \sqrt{U_\mu U^\mu} \mathcal{H}_q, \label{Hqb}
\end{align}
respectively.

By using Eqs.~(\ref{Hqa}) or~(\ref{Hqb}) one can cast the total covariant Lagrangian for a free particle, 
\begin{align}
    L &= \sqrt{U_\mu U^\mu} \bigl( \mathcal{L}_{ext} + \mathcal{L}_{int} \bigl),
\end{align}
as
\begin{align}\label{LH}
    L &= -\sqrt{U_\mu U^\mu} \bigl( m + \mathcal{H}_{q} - P_q q' \bigl) \nonumber \\
    &= -\sqrt{U_\mu U^\mu} \biggl( m + \mathcal{H}_q - \frac{P_q \dot{q}}{\sqrt{U_\mu U^\mu}} \biggl).
\end{align}
Now, we can use Eq.~(\ref{LH}) to compute the external conjugate 4-momentum  $\mathbb{P}_\mu \equiv {\partial L}/{\partial U^\mu}$ yielding 

\begin{align}
    \mathbb{P}_\mu  = - (m + \mathcal{H}_q) \frac{U_\mu}{\sqrt{U_\alpha U^\alpha}}\equiv -P_\mu.
\end{align}
Thus, we see immediately that the internal energy contributes to the mass in the dispersion relation, as should be the case (note that this contribution comes out automatically from covariance of the Lagrangian under reparametrization). We thus define the total dynamical mass $M$ as

\begin{align}\label{M2}
     M^2\equiv P_\mu P^\mu \approx (m+\mathcal{H}_q)^2. 
\end{align}

The total Hamiltonian constraint, for $M \neq 0$, can now be written as

\begin{align}
    H_{\rm free} &\equiv \mathbb{P}_\mu U^\mu + P_q \dot{q} - L \nonumber \\
    &= \sqrt{U_\alpha U^\alpha} \biggl(- \frac{P_\mu P^\mu}{M} + M \biggl) \approx 0.
\end{align}
We note that we can decompose this constraint as
\begin{align}
    H_{\rm free} &= \frac{\sqrt{U_\alpha U^\alpha}}{M}\left(M^2 - P^2\right) \nonumber \\
    &=\frac{\sqrt{U_\alpha U^\alpha}}{M} \bigl( M + \sqrt{P^2} \bigl)\bigl( M - \sqrt{P^2} \bigl) 
\end{align}
where $P^2 \equiv P_\mu P^\mu$. Be defining 
\begin{eqnarray}
  H_{\rm free}^{(+)}&\equiv  \bigl( M + \sqrt{P^2} \bigl), \nonumber \\
  H_{\rm free}^{(-)}&\equiv  \bigl( M - \sqrt{P^2} \bigl), 
\end{eqnarray}
we can cast the Hamiltonian constraint as
\begin{equation}
H_{\rm free}\equiv \frac{\sqrt{U_\alpha U^\alpha}}{M} H_{\rm free}^{(+)} H_{\rm free}^{(-)} \approx 0. \label{Hq free decomposition}
\end{equation}

If our internal Hamiltonian $\mathcal{H}_q$ is such that $M = m + \mathcal{H}_q > 0$, we must have that $H_{\rm free}^{(+)}>0$. As a result, all physical configurations obeying the Hamiltonian constraint must have a vanishing $H_{\rm free}^{(-)}$, i.e., $H_{\rm free} \approx 0 \Leftrightarrow H_{\rm free}^{(-)} \approx 0$. (Reciprocally, if $\mathcal{H}_q$ was sufficiently negative such that $M<0$, we would have that $H_{\rm free}^{(-)}<0$ and $H_{\rm free} \approx 0 \Leftrightarrow H_{\rm free}^{(+)} \approx 0$.)

\subsection{Coupling with external fields} \label{covariant couplings}

The next feature that we want to incorporate in our model are (covariant) dynamical mechanisms for accelerating our clocks. These will be crucial in the quantized formalism since, unlike the semiclassical case, one cannot ascribe a trajectory to a particle with a quantized center of mass. Any prescription for the motion of center-of-mass DoFs must be done properly at a dynamical level. The corresponding dynamics will then bear nontrivial consequences for the relation between the internal and external degrees of freedom, as we shall see. Employing such a covariant formulation jointly with (suitably chosen) external interactions to dynamically accelerate the particles is one of the novel contributions of our work. In this context, one can regard the external fields as playing an analog role to external potentials in non-relativistic quantum mechanics. However, the covariance requirements for our action drastically constrains the kinds of couplings that we are allowed to add in the theory. This, in turn, will have implications for how internal and external DoFs couple. This will help to clarify from first principle whether or not one can recover an unitary time evolution for accelerating (or interacting clocks), an issue that is discussed in the nonrelativistic limit in~\cite{aharonov}.

Let us start with the simplest possible case, a scalar field $\phi(x)$. In addition to the covariance already discussed, we will require our coupling to be local, i.e., it must only depend on the field's (and possibly on its derivatives) value at the particle's position $x$. For simplicity, we shall choose a coupling which is linear in the field amplitudes in which case the total Lagrangian will read

\begin{align}
    L_{SF}(x^\mu,U^\mu) &= \sqrt{U_\mu U^\mu} \Bigl[ m + \mathcal{H}_q - \frac{P_q \dot{q}}{\sqrt{U_\mu U^\mu}} + g\phi(x) \Bigl], \label{lag SF}
\end{align}
 where $g$ is a coupling constant. If we write the Euler-Lagrange equations for the CM motion using the proper time, $s=\tau$, we find

\begin{align}
\bigl[M+ g\phi(x) \bigl]a_\mu = - g  \, h_\mu^{\;\;\nu} \partial_\nu \phi(x) \label{EL_SP+SF},
\end{align}
where $h_\mu^{\;\;\nu} \equiv \delta_\mu^{\;\;\nu} - u_\mu u^\nu$ is the orthogonal projector with respect to the particle's 4-velocity. We note that $a_\mu$ satisfies the constraint given in Eq.~\eqref{4-normalization}. We can see from Eq.~\eqref{EL_SP+SF} that the field's role in the dynamics is twofold: {\bf (i)} its gradient produces a 4-force in the right-hand side (RHS) of and  {\bf (ii)}  its  amplitude gives a contribution to the particle's inertia appearing in the left-hand side (LHS). Thus, for the coupled particle, we define a position-dependent dynamical mass as

\begin{align}\label{Mx}
    \mathcal{M}(x) = M + g\phi(x). 
\end{align}
This dynamical mass is precisely the one appearing in the dispersion relation in the constrained phase space

\begin{align}
   \mathbb{P}_\mu = \frac{-\mathcal{M}(x)U_\mu}{\sqrt{U_\alpha U^\alpha}}\equiv -P_\mu \qquad \Rightarrow \qquad P_\mu P^\mu \approx \mathcal{M}^2(x).
\end{align}

By explicitly computing the Hamiltonian $ H = \mathbb{P}_\mu U^\mu + P_q \dot{q} - L $, it is easy to show that it can be decomposed into positive and negative terms [as in Eq.~ \eqref{Hq free decomposition}] yielding the Hamiltonian constrain 

\begin{align}
    H_{SF} &\propto \mathcal{M}^2(x) - P^2 \nonumber \\
    &= \bigl( \mathcal{M}(x) + \sqrt{P^2} \bigl)\bigl(\mathcal{M}(x) - \sqrt{P^2} \bigl) \nonumber \\
    &\equiv H_{SF}^{(+)} H_{SF}^{(-)} \approx 0. \label{Hq scalar decomposition}
\end{align}
If $\mathcal{M}(x)>0$ we will have that $H_{SF} \approx 0 \Leftrightarrow H_{SF}^{(-)} \approx 0$.

Another interesting case, which has a more direct physical interest,  is the one of a charged particle coupled to the electromagnetic (EM) field.  In this case, the simplest choice of action complying with covariance and locality requirement is given by
\begin{align}
    L_{EM} = - \sqrt{U_\mu U^\mu} \Bigl[ m + \mathcal{H}_q - \frac{P_q \dot{q}}{\sqrt{U_\mu U^\mu}}  \Bigl] - \, qU_\mu A^\mu(x),
\end{align}
with $A^\mu(x)$ being the 4-vector potential. The dynamical equations for the CM Dofs take the familiar covariant form
\begin{align}
    M a_\mu = qF_{\mu\nu}u^\nu, \qquad F_{\mu\nu} \equiv \partial_\mu A_\nu - \partial_\nu A_\mu,
\end{align}
which, again, satisfy the constraint~\eqref{4-normalization}. Here, however, the field only contributes to the RHS of the equations, i.e., to the force, but not to the mass on the LHS. The conjugate 4-momentum in this case reads

\begin{align}
    \mathbb{P}_\mu = -\frac{MU_\mu}{\sqrt{U_\alpha U^\alpha}} -qA_\mu \equiv -P_\mu,
\end{align}
yielding the constraint

\begin{align}
    (P_\mu - qA_\mu(x))^2 \approx M^2.
\end{align}

Like the scalar and free cases, the Hamiltonian constraint can be decomposed as

\begin{align}
    H_{EM} &\propto M^2 - \Pi^2 \nonumber \\
    &= \bigl( M + \sqrt{\Pi^2} \bigl)\bigl( M - \sqrt{\Pi^2} \bigl) \nonumber \\
    &\equiv H_{EM}^{(+)} H_{EM}^{(-)} \approx 0. \label{Hq EM decomposition}
\end{align}
where, for convenience, we have defined  $\Pi_\mu \equiv P_\mu - qA_\mu(x)$. Again, for $M>0$, we will have that $H_{EM} \approx 0 \Leftrightarrow H_{EM}^{(-)} \approx 0$.


\section{Quantization}\label{sec:Quantization}

Having laid down the classical framework for our covariant description of relativistic clocks in terms of constrained systems, we dedicate this section to its quantization. Constrained systems and techniques for their quantization are ubiquitous in modern physics \cite{teitelboim, QCS2, QCS3, QCS4, QCS5}. The framework presented here is inspired by \cite{flaminiaSTQRFs} and \cite{smithQC1} but with several  fundamental differences. Differently from \cite{flaminiaSTQRFs}, we do not introduce global gauge constraints to produce a perspective-neutral structure\footnote{Heuristically, we would argue that such a structure could only make sense if we were trying to describe a `toy model universe', with no systems external to it that could potentially interact with our clocks to, e.g., infer their positions. Our description, on the contrary, aims to model only a small system which can presumably interact with external systems, both to influence their dynamics and to make measurements on it.} and we introduce the clocks' internal Hamiltonian covariantly, such that the internal energy associated to it appears in the clocks' dispersion relation. Differently from \cite{smithQC1}, we consider time evolution directly with respect to the internal proper times of the clocks (rather than with respect to external coordinate time variables), and we evaluate time dilation with a fully quantum conditional density operator $\rho_{\tau_2|\tau_1}$, rather then conditional probabilities on joint observables $P(T_2=\tau_2|T_1=\tau_1)$. Differently from both,  we do not employ ideal clocks and we decompose our quadratic Hamiltonian constraints in positive and negative mass sectors (rather than energy sectors).


\subsection{Quantum clocks and temporal POVMs}\label{ccPs}

Before looking at the quantization of our fully constrained system, we want to analyze more closely the subsystem whose observables will allow us to measure time, which we will refer as  the `quantum clock'. We denote the Hilbert space of the clock DoF as $\mathcal{H}_C$ and its internal Hamiltonian as $H_C$, such that the states $\ket{\varphi}_C \in \mathcal{H}_C$ evolve in proper time in the usual manner

\begin{align}
    i \frac{d}{d\tau} \ket{\varphi}_C = H_C \ket{\varphi}_C.
\end{align}
In general, we would like to associate the quantized DoF in $\mathcal{H}_C$ with the classical internal dynamical variable $q$ (where, in the notation used in Sec.~\ref{sect:CovPM}$, (q,P_q) \in \mathcal{P}_q$) and $H_C$ with the classical internal Hamiltonian $\mathcal{H}_q$ via the usual quantization procedure. While this will indeed be done when  quantizing classical clocks, it is worth noting that not all quantum clocks will have a classical analog (as will be the case for clocks with a finite-dimensional Hilbert space, such as spin systems).


Essentially, the two relevant properties of our clocks will be the structure of their Hilbert spaces and the spectrum $\sigma(H_C)$ of their Hamiltonians. By far, the most used model for quantum clocks in the literature are ideal clocks, whose Hilbert space $\mathcal{H}_C = \mathcal{L}^2(\mathbb{R})$ is analogous to that of a 1-dimensional wave-function, and whose Hamiltonian has a continuous unbounded spectrum $\sigma(H_C)=\mathbb{R}$. These special features allow one to define a self-adjoint observable $T$ conjugate to $H_C$, $[T,H_C]=i$, much like a position-momentum pair. The eigenstates $\ket{t}$ of $T$, $T\ket{t}=t\ket{t}$, will satisfy the usual orthogonality and completeness properties

\begin{align}
    \braket{t'|t} = \delta(t-t'), \qquad \qquad \int\limits_{-\infty}^{\infty} dt \,\ketbra{t}{t} = \mathlarger{\mathbb{1}},
\end{align}
and will transform covariantly with the internal Hamiltonian $H_C$

\begin{align}
    e^{-iH_C \,s}\ket{t} = \ket{t+s}.
\end{align}

As discussed in the introduction, although ideal clocks are an useful idealization to model several aspects of quantum clocks in nonrelativistic scenarios, they have many unphysical aspects. Fortunately, much progress has already been made with regard to the description of physical clocks. By using POVMs to define time observables, one is able to consider much more realistic internal Hilbert spaces and internal Hamiltonians~\cite{trinity, trinity2, flaminiaSTQRFs, smithQC1}. Hence, due to its fundamental role, we shall briefly review the POVM construction here. We will follow closely the notation of \cite{trinity}.

First, we define our clock's proper-time states $\ket{\tau}_C$ as those who transform covariantly with respect to the internal Hamiltonian $H_C$, i.e., 

\begin{align}
    \ket{\tau+\tau'}_C = e^{-i H_C \mathlarger{\tau'}} \ket{\tau}_C,
\end{align}
Such states can be easily constructed from the energy eigenstates $\ket{\epsilon}_C$, with $H_C\ket{\epsilon}_C = \epsilon\ket{\epsilon}_C$ and $\epsilon\in \sigma(H_C)$, as 

\begin{align}
    \ket{\tau}_C = \int\limits_{\sigma(H_C)} d\mu(\epsilon) \ket{\epsilon}_{\!C}\, e^{i g(\epsilon)} e^{- i \epsilon \tau}, \label{ket tau_C}
\end{align}
where the integral is carried over the spectrum $\sigma(H_C)$ with measure $\mu(\epsilon)$\footnote{Note that this notation includes both continuous and discrete spectra.}. The inner product between two of those states is then

\begin{align}
    \braket{\tau'|\tau} = \int\limits_{\sigma(H_C)} d\mu(\epsilon) e^{-i\epsilon(\tau-\tau')} \equiv \Delta(\tau-\tau'),
\end{align}
with exact form for $\Delta(\tau-\tau')$ will depend on the spectrum of the clock. For ideal clocks, $\sigma(H_C)=\mathbb{R}$ and $d\mu(\epsilon)=d\epsilon/\sqrt{2\pi}$, so that we recover $\Delta(\tau-\tau')=\delta(\tau-\tau')$; but generally this inner product is nonvanishing for $\tau\neq\tau'$. It is important to note that the choice of clock states is not unique, one has the freedom to choose an arbitrary set of relative phases $g(\epsilon)$. 

Having chosen the clock states $\ket{\tau}_{\!C}$, we define the  corresponding temporal POVM $T$ (as well as its statistical moments $T^{(n)},$ $n\in \mathbb{Z}^+$) on $\mathcal{H}_C$ as
\begin{align}
    T^{(n)} \equiv  \nu\int_G d\tau \, \tau^n \ket{\tau}\!\bra{\tau}, \qquad n \in \mathbb{N}, \label{T^(n)}
\end{align}
where $\nu$ is a normalization constant and $G$ is the 1-parameter group of all possible distinguishable values of $\tau$. For ideal clocks we have $G = \mathbb{R}$, meaning the clock runs from $-\infty$ to $+\infty$, while for clocks with a regular discrete spectrum $G$ is compact and thus, the internal states run over periodic cycles. We fix $\nu$ by imposing that the zeroth moment $T^{(0)}$ sums up to the identity $\mathlarger{\mathbb{1}}_C$ in $\mathcal{H}_C$, {expressing the completeness relation of our POVM. The completeness relation allows us to write any clock state $\ket{\varphi}_C$ in a proper time basis as

\begin{align}
    \ket{\varphi}_{\!C} = \nu\int d\tau \ket{\tau}\!\braket{\tau|\varphi} \equiv \int d\tau \ket{\tau}_{\!C}\varphi(\tau).
\end{align}

If the clock is in a given state $\rho_C$, the probability that we measure its proper time $\tau$ in a certain interval $ I \subset G$ is simply

\begin{align}
    P_I(\rho_C) = \nu\int_I d\tau   \Tr\bigl\{ \rho_C \ketbra{\tau}{\tau}  \bigl\}.
\end{align}
As can be straightforwardly verified from Eqs.~\eqref{ket tau_C} and~\eqref{T^(n)}, the first moment, $T \equiv T^{(1)}$, will be a Hermitian operator canonically conjugated to $H_C$, i.e., $[T,H_C] = i$. However, in general, $T$ will not be self-adjoint and, thus, the states $\ket{\tau}_{\!C}$ will generally not be eigenstates of $T$ nor will they be orthogonal
\begin{align}
    \braket{\tau'|\tau} \equiv \Delta(\tau'-\tau) \neq \delta(\tau'-\tau).
\end{align}
In fact, the finite time resolution that appears for physical clocks is a feature of our theory and they yield a quantum `fuzziness' for temporal measurements. 

\subsection{Dirac quantization, physical Hilbert Space and Hamiltonian decomposition} \label{dirac quantization}

Let us now describe the quantization of our systems of interest (see Sec.~\ref{sect:CovPM}). For more details on the quantization of constrained systems, we refer the reader to \cite{teitelboim, QCS2,QCS3,QCS4,QCS5} (as well as \cite{trinity,trinity2} for its application in the context of quantum clocks).

We shall use here the Dirac quantization procedure, where one first quantizes the full theory and then handles the gauge symmetry and its corresponding constraints at a quantum level. Hence, one starts with a kinematical Hilbert Space, $\mathcal{H}_{kin}$, built from the kinematical Phase Space $\mathcal{P}_{kin}$, on which we implement the covariant commutation relations $[X^\mu,\mathbb{P}_\nu] = i\delta^\mu_{\;\,\nu}$.
In this ancillary space, the classical constraints $\{\mathcal{C}_I: \mathcal{P}_{kin} \rightarrow \mathbb{R}\}_{I=1,...N}$ become linear operators on $\mathcal{H}_{kin}$, $\{C_I(X,P)\}_{I=1,...N}$. For first-class constraints, i.e., constraints that commute with each other weakly~\cite{teitelboim}, one defines the physical Hilbert Space $\mathcal{H}_{phys} \subset \mathcal{H}_{kin}$ as the sub-space of states $\ket{\Psi}\!\rangle$ that satisfy all constraint equations\footnote{There is a number of subtleties in those steps: {\bf (1)} it is necessary that all the constraints commute at least weakly (classically, their Poisson brackets must vanish in $\mathcal{P}_{phys}$, $\{\mathcal{C}_I,\mathcal{C}_J\} \approx 0$). If that is not the case, one cannot build $\mathcal{H}_{phys}$ in a straightford manner as a linear subspace and must resort to the use of equivalence relations to identify physically equivalent states in $\mathcal{H}_{kin}$. {\bf (2)} If any of the constraints $C_I$ has a continuous spectrum around 0, then their corresponding physical states $\ket{\Psi}\!\rangle$ (satisfying $C_I\ket{\Psi}\!\rangle=0$) will be improper states in $\mathcal{H}_{kin}$. In this case one must define suitable Hilbert Space completions and redefine the inner product in $\mathcal{H}_{phys}$~\cite{trinity, trinity2}.} 

\begin{align}
    C_I\ket{\Psi}\!\rangle = 0, \; \forall I=1,\cdots,N.
\end{align}

In our systems of interest, we start from $N$ independent classical Hamiltonian constraints $\mathcal{H}_I \approx 0$ which commute everywhere, $\{\mathcal{H}_I, \mathcal{H}_J\} = 0$. Thus, they will give rise to $N$ independent quantum constraint equations

\begin{align}
    H_I\ket{\Psi}\!\rangle = 0, \; \forall I=1,\cdots,N,
\end{align}
with all the $H_I$ commuting in the whole $\mathcal{H}_{kin}$, i.e., $[H_I,H_J]=0$ for all $I,J=1,\cdots N$. For simplicity, let us first focus on the case of a single clock (and thus of a single Hamiltonian constraint $H$). In general, we can decompose our kinematic Hilbert space in a tensor product  $\mathcal{H}_{kin} = \mathcal{H}_{C} \otimes \mathcal{H}_{S}$ between (internal) clock DoFs and the rest of the system. Whenever we can break a Hamiltonian constraint as 
\begin{align}
    H &= H_C + H_S \nonumber \\
    &\equiv  H_C \otimes \mathbb{1}_S + \mathbb{1}_C \otimes H_S, \label{HC+HS}
\end{align}
our physical states $\ket{\Psi}\rangle \in \mathcal{H}_{phys}$ (which we also refer to as \textit{history states}) can be written in the following entangled form
\begin{align}\label{psiphystau}
    \ket{\Psi}\rangle = \int d\tau \ket{\tau}_C \ket{\psi(\tau)}_S,
\end{align}
with the system's partial state satisfying an ordinary Schrödinger equation with respect to the clock's internal time
\begin{align}
    i \frac{d}{d\tau} \ket{\psi(\tau)}_S = H_S \ket{\psi(\tau)}_S.
\end{align}

Of course, in general, $H$ will not have the simple separable form given in Eq.~\eqref{HC+HS}\footnote{In fact, an essential feature of gravity is that it couples universally to everything and, thus, no system can remain truly isolated in this context. This hints that a fundamental feature of quantum gravity is that this separability must be broken, being valid at best as an approximation.}. When separability is broken, the simple picture above changes dramatically: the partial state $\ket{\psi(\tau)}_S$ evolves not via an ordinary time-local, first-order, Hermitian Schrödinger equation, but generally via time non-local and non-unitary equations \cite{aharonov,IQC}. Here, however, we shall focus on the simpler regime where the Hamiltonian separability holds (at least in a good approximation), but where there are still relevant quantum fluctuations for the physical observables keeping track of proper time.



We can see from Eqs.~(\ref{Hq free decomposition}),~(\ref{Hq scalar decomposition}), and~(\ref{Hq EM decomposition}) that as long as the dynamic masses~\eqref{M2} or~\eqref{Mx} remained positive, imposing the Hamiltonian constraint is equivalent to imposing $H^{(-)}\approx 0$. As $H^{(-)}$ is given by 
\begin{eqnarray}
  H^{(-)}&=& M - \sqrt{P^2},   \label{H-} \\
 H^{(-)}_{SF}&=& \mathcal{M}(x) - \sqrt{P^2},   \label{H-SF} \\
\end{eqnarray}
and 
\begin{eqnarray}
 H^{(-)}_{EM}= M - \sqrt{\Pi^2},   \label{H-EM} 
\end{eqnarray}
for the free, scalar, and electromagnetic coupled cases  respectively (where we recall that $\Pi\equiv P_\mu - eA_\mu$), we say that the solutions of $H_\zeta^{(-)}$, $\zeta={\rm free}, SF, EM,$ lie in the {\em positive mass sector}. In fact, as long as $\mathcal{H}_q$ (and $\phi(x)$) are bounded, such that $\mathcal{H}_q+g\phi(x)>-m$, all (classical) physical states lie in the positive mass sector. For this reason, in the  quantized theories for the free and electromagnetic coupled cases, we shall always demand that the spectrum of $H_C$ is bounded from below such that ${\rm inf}\;\sigma(H_C)>-m \, \Leftrightarrow \, \sigma(M)>0$. (The coupling with a scalar field, which results in a position-dependent mass, brings further complications that we mention ahead.) As a result, 
\begin{align}
    H_\zeta \Pket = 0 \qquad \leftrightarrow \qquad H_\zeta^{(-)}\Pket = 0, \; \zeta={\rm free}, EM,
\end{align}
and thus $\Pket \in \mathcal{H}_{phys}$ if and only if $\Pket \in \ker(H^{(-)})$.

We must note, however, that for the quantum case there is a subtlety. If $H_\zeta^{(+)}$ and $H_\zeta^{(-)}$ do not commute, then in general it will not be true that $H_\zeta=H_\zeta^{(+)}H_\zeta^{(-)}$ as we will have an ordering ambiguity in defining the quantized Hamiltonian. For the  models that we considered in the previous section, we have that $[H_\zeta^{(+)},H_\zeta^{(-)}]=0$ for the free and electromagnetically coupled particles but $[H_{SF}^{(+)},H_{SF}^{(-)}]\neq0$ for particles coupled with a scalar field. This comes from the fact that $[\mathcal{M}(X),\sqrt{P^2}]\neq 0$ in such a case\footnote{A similar issue happens to particles coupled to an external gravitational field, as $g_{\mu\nu}(X)P^\mu P^\nu$ also has ordering ambiguities on quantization.}. Thus, to avoid such a difficulty, we will restrict ourselves to the free and electromagnetic case in what follows, leaving some questions regarding the scalar-coupled particle for future work.

Here, we emphasize that the decomposition we are using is distinct from the one usually adopted in the literature. In \cite{smithQC1,flaminiaSTQRFs}, for instance, one picks a preferred coordinate system to cast $p^2 =g_{\mu\nu}p^\mu p^\nu$ as $(p^0)^2 - \mathbf{p}^2 \equiv  (p^0)^2 + g_{ij}p^ip^j$ and then decomposes the quadratic Hamiltonian into positive and negative energy sectors, $p^0>0$ and $p^0<0$, respectively. Then, in \cite{smithQC1}, the authors  use the positive-energy Hamiltonian (linear in $p^0$) to compute the external (coordinate) time evolution for free clocks. In \cite{flaminiaSTQRFs}, the author adds linear terms (linear in the clock Hamiltonian) in an extra global gauge constraint (which is used to enforce a perspective-neutral superstructure). This is used to compute internal time evolution with respect to a fiducial clock's proper time\footnote{As the author considers particles subject to a gravitational field, ordering ambiguities arise and the factorization of the theory in positive- and negative-energy sectors also becomes nontrivial. There, these problems are handled by restricting the model to a weak field regime, where the terms arising from nontrivial commutators become negligible in leading order.}. Both these approaches, although  substantially different from one another, manifestly break covariance.


In this work, we propose a somewhat hybrid approach, while maintaining the Hamiltonian decomposition manifestly covariant. For the sake of clarity, let us first illustrate this decomposition in the case of a free clock. Our linear constraint $H_{\rm free }^{(-)}$ yields the constraint equation

\begin{align}
    \bigl( H_C + m - \sqrt{P^2} \,\bigl)\Pket = 0,
\end{align}
resulting in the conditional proper-time evolution for the external DoFs

\begin{align}
    i \frac{d}{d\tau} \ket{\psi(\tau)}_S = \bigl(m-\sqrt{P^2} \bigl)\ket{\psi(\tau)}_S \equiv H_S\ket{\psi(\tau)}_S.
\end{align}
Then, we can define an initial condition for $\ket{\psi(\tau)}_S$ for $\tau=0$. Using the 4-momentum eigenstates $\ket{p}_S \in \mathcal{H}_S$, $P^\mu\ket{p}_S=p^\mu\ket{p}_S$, we write

\begin{align}
    \ket{\psi(\tau\!=\!0)}_S = \int \frac{d^4p}{(2\pi)^\frac{4}{2}} \psi_0(p^\mu) \ket{p}_S. \label{initial condition}
\end{align}

We note, however, that the values for the coefficients $\psi_0(p^\mu)$ are not completely arbitrary. To enforce the Hamiltonian constraint, they can only be nonvanishing for $p^2 \in \sigma(M^2),\;  M=H_C+m$. Thus, $\psi_0(p)$ has support in a subset inside the (momentum) light cone, (i.e., for timelike $p^\mu$). This supporting region includes both future-oriented (positive-energy, $p^0>0$) and past-oriented (negative-energy, $p^0<0$) values for $p^\mu$, see Fig.~ \ref{supp(psi_0)}). Thus, the choice of which energy sectors appear in our physical states translates into a choice of initial condition. If, for instance, we are only interested in positive-energy solutions, we must pick a $\psi_0$ which is nonvanishing in the $p^0>0$ portion of the support. Having chosen and initial condition, we can then evolve the initial state in Eq. \eqref{initial condition} with the Hamiltonian $H_S$ yielding

\begin{align}  \label{evolved stateP}
    \ket{\psi(\tau)}_S &= e^{-iH_S\mathlarger{\tau}} \ket{\psi(\tau=0)}_S \nonumber \\
    &= \int \frac{d^4p}{(2\pi)^\frac{4}{2}} \psi(p^\mu;\tau)\, \ket{p}_S 
\end{align}
where $\psi(p^\mu;\tau)\equiv \psi_0(p^\mu) e^{-i(m-\sqrt{p^2})\tau}.$ Alternatively, by using ${}_{S}\braket{x|p}_S=(2\pi)^{-\frac{4}{2}}e^{-ip_\mu x^\mu}$, one can write $ \ket{\psi(\tau)}_S$ in terms of position eigenstates as   
    \begin{align}
         \ket{\psi(\tau)}_S  = \int \frac{d^4x}{(2\pi)^\frac{4}{2}} \psi(x^\mu;\tau)\, \ket{x}_S  \label{evolved state}
\end{align}
with
\begin{equation}
    \psi(x^\mu; \tau)\equiv \int \frac{d^4p}{(2\pi)^\frac{4}{2}} \psi_0(p^\mu) e^{-i(m-\sqrt{p^2})\tau}e^{-ip_\mu x^\mu}, \label{evolved state 2}
\end{equation}
allowing us to directly analyze the phenomenology of our solutions in spacetime terms.
\begin{figure}[H]
    \centering 
    \includegraphics[width=0.9\linewidth]{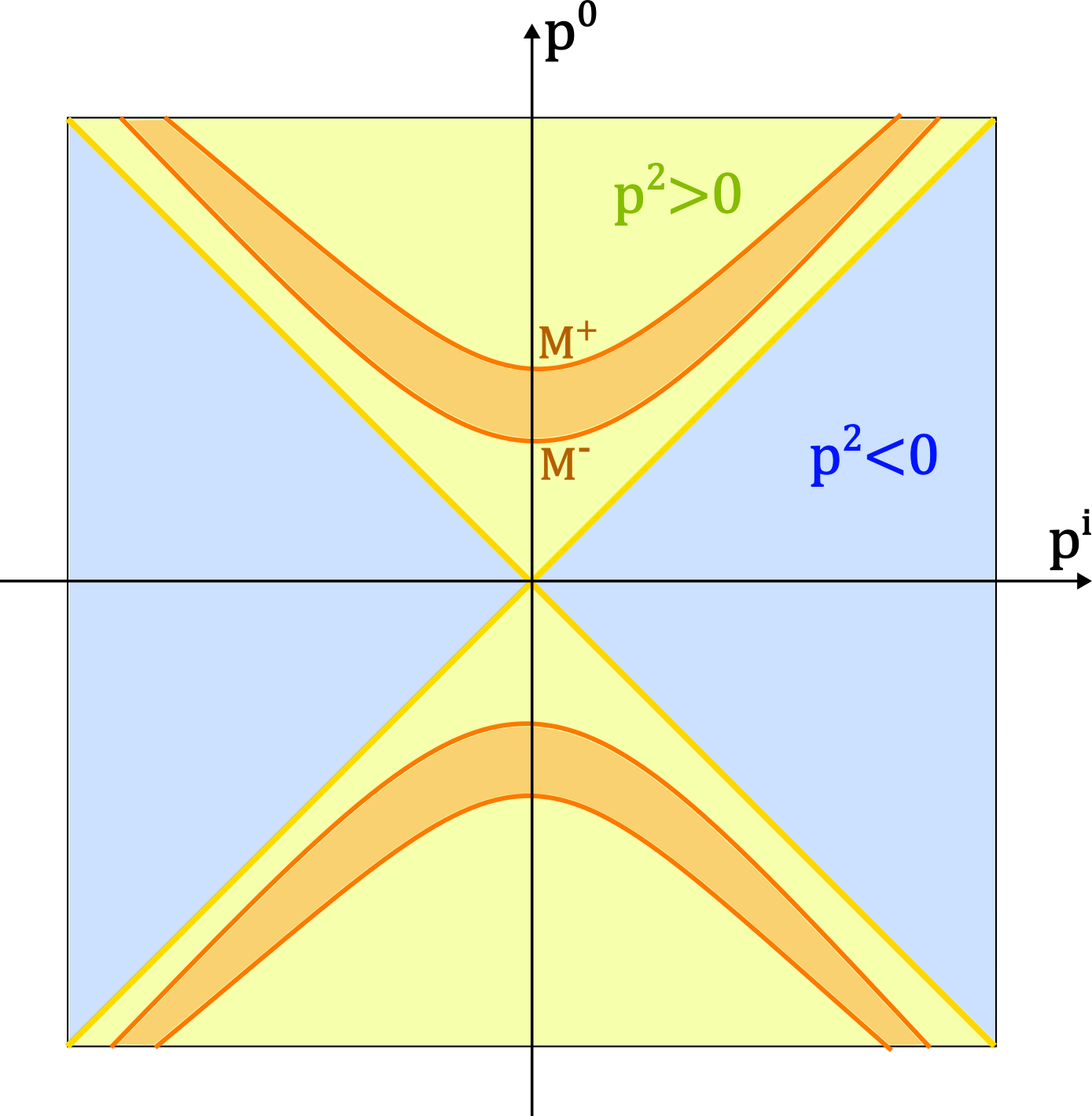}
    \caption{Support of $\psi_0$ in momentum space, represented by the orange stripes in the future and past light cones. This region is delimited by $(M^-)^2<p^2<(M^+)^2$, where $M^\pm = m + \epsilon^\pm$, and $\epsilon^+$ ($\epsilon^-$) is the maximum (minimum) value in the internal energy spectrum $\sigma(H_C)$ of the clock.}
    \label{supp(psi_0)}
\end{figure}
\noindent By using Eq.~(\ref{evolved stateP}) [or Eq.~\eqref{evolved state}] in Eq.~\eqref{psiphystau} one can write the physical state of the total (Clock+Particle) system.  

A complementary way to write down the physical states which automatically enforces all the constraints is the following. Let $\ket{\Psi}\rangle$ be a physical state satisfying the constraint $H_{\rm free}^{(-)}$. By using the energy eigenstates $\ket{\epsilon}_C$ of $H_C$ and momemtum eigeinstates $\ket{p}_S$, it can be written as

\begin{align}
    \ket{\Psi}\!\rangle = \int d\mu(\epsilon) \frac{d^4p}{(2\pi)^{4/2}} \, \ket{\epsilon}_C \ket{p}_S \Psi(p^\mu;\epsilon) \delta\bigl(m+\epsilon-\sqrt{p^2}\bigl), \label{free PSI}
\end{align}
where $\Psi(\epsilon;p^\mu)$ is an arbitrary complex amplitude. By using the decomposition of the identity $\nu \int d\tau \ket{\tau}\!\bra{\tau}=\mathlarger{\mathbb{1}}_C$ together with $\braket{\tau|\epsilon}=e^{i\epsilon\tau} e^{-ig(\epsilon)}$ [see Eq.~\eqref{ket tau_C}], we can cast $\Pket$ as

\begin{align} \label{alternative evolved state}
     \ket{\Psi}\!\rangle 
     &= \int d\tau \ket{\tau}_{C} \Biggl[\nu  \int d\mu(\epsilon) \frac{d^4p}{(2\pi)^{4/2}} \, \ket{p}_{S}\, \nonumber \\
     & \qquad\quad \times e^{i\epsilon\tau} e^{-ig(\epsilon)} \Psi(p^\mu;\epsilon) \delta\bigl(m+\epsilon- \sqrt{p^2}\bigl) \Biggl],
\end{align}
By comparing Eqs.~\eqref{evolved stateP}  and~\eqref{alternative evolved state} we can write
\begin{align}
    \psi(p^\mu;\tau) = \nu\!\int d\mu(\epsilon) \, e^{i\epsilon\tau} e^{-ig(\epsilon)} \Psi(p^\mu;\epsilon) \delta(m+\epsilon-\sqrt{p^2}).
\end{align}
We have now two equivalent ways to postulate initial conditions for our system, either in the reduced states $\ket{\psi_0}_S$ or directly at the kinematic states $\Psi(p^\mu;\epsilon)$.  The latter has the advantage of automatically enforcing the constraint (and being more ``global''), whereas the former is more intuitive to interpret in terms of spacetime observables, with the conditional states $\ket{\psi(\tau)}_S$ being closer to the ones used in usual quantum mechanics.

The formal construction of $\Pket$ and $\ket{\psi}_{\!S}$ can be carried in an entirely analogous fashion for a more general $H_S$. If we choose a basis for $\mathcal{H}_S$ with eigenstates of the form $\ket{E,\lambda_E}_S$, where $E \in \sigma(H_S)$ and $\lambda_E$ represent extra quantum numbers (accounting for possible degeneracies in the energy levels), we have
\begin{align}
    \ket{\Psi}\!\rangle = \int d\mu(\epsilon, E,\lambda_E)  \, \ket{\epsilon}_{\!C} \ket{E,\lambda_E}_{\!S} \Psi(E,\lambda_E;\epsilon) \delta\bigl(\epsilon- E\bigl) \label{general PSI}
\end{align}
and
\begin{align}
    \psi(E,\lambda_E;\epsilon) = \nu \!\!\int d\mu_C(\epsilon) e^{i\epsilon\tau} e^{-ig(\epsilon)} \Psi(E,\lambda_E;\epsilon)\delta(\epsilon-E),
\end{align}
with $ d\mu(\epsilon, E,\lambda_E)\equiv d\mu_C(\epsilon) \, d\mu_S(E) d\mu_E(\lambda_E).$

To end this section, let us point out how our formalism enables one to consider the (temporal) quantum reference of an accelerated particle and still obtain an unitary evolution for its partial state, $|\psi(\tau)\rangle_S$, describing its external DoFs. 

Indeed, consider, e.g., a charged particle with mass $m$ and charge $q$ accelerated by some electromagnetic field $A^\mu$. Then, as we discussed earlier in this section, whenever $\sigma(M)>0$, where $M=m+H_C$, imposing the EM Hamiltonian constraint is equivalent to impose the constraint 
\begin{equation}
  \bigl( H_C + m - \sqrt{\Pi^2} \,\bigl)\Pket = 0,
\end{equation}
where $\Pi_\mu=P_\mu-qA_\mu$. This, in turn, will yield the unitary evolution
\begin{align}
    i \frac{d}{d\tau} \ket{\psi(\tau)}_S = \bigl(m-\sqrt{\Pi^2} \bigl)\ket{\psi(\tau)}_S
\end{align}
for the partial states $|\psi(t)\rangle_S$ describing the CM DoFs. 

Hence, when considering a full covariant treatment, one obtain a unitary evolution for non-inertial quantum clock frames, in contrast to previous analysis based on adding relativistic corrections to the non-relativistic equations \cite{aharonov}.

\subsection{Single clock: clock states and local time evolution}

Now that we have derived a general form for the physical states of our theory with a single linear Hamiltonian constraint, let us analyze more closely the time evolution in the paradigmatic case of a free clock. This will serve to illustrate the main ideas of our formalism in a simple manner. 

Let us first comment on the choice of ``good'' initial state $\psi_0(p^\mu)$ [or $\Psi(p^\mu;\epsilon)$] in this spacetime-covariant constrained theory. Among the simplest possible solutions are plane-waves, $\psi_0(p^\mu) \propto\delta(p^\mu-p_*^\mu)$, which have time-evolution are given by

\begin{align}
    \ket{\psi(\tau)} \propto \ket{p_*}e^{-i\left(m-\sqrt{p_*^2}\right)\tau}.
\end{align}
However, the above equation implies that there is no correlation between proper time and external DoF (coordinate time or position). Indeed, this solution is separable $\Pket \propto\ket{\epsilon_*}_{\!C} \ket{p_*}_{\!S}$, where $\epsilon_*=\sqrt{p_*^2}-m$ [the situation is similar for combination of plane waves in a well-defined energy subspace $\ket{\psi_0}_S\propto \delta\bigl( \sqrt{p^2}-m-\epsilon_*\bigl)$].  Hence, for our purposes, they are trivial and uninteresting. Thus, to construct ``good'' initial states, one needs  entanglement between internal and external DoFs, superposing different energy levels with comparable amplitudes\footnote{The intuition is similar to the construction of the internal proper-time states $\ket{\tau}_{\!C}$ alone. For these states to be maximally distinguishable in time, they must be maximally spread in energy, being constructed with equal amplitudes across the entire clock spectrum $\sigma(H_C)$.}.

We shall consider clocks with a continuous bounded spectrum $\sigma(H_C)=[-\Omega,\Omega]$ ($\Omega<m$) and Gaussian clock states

\begin{align}
    \psi_0(p) = N e^{-\frac{(p_0-m)^2}{2\Delta_0^2}}e^{-\frac{\mathbf{p}^2}{2\Delta^2}} \,\chi(\sqrt{p^2}-m), \label{gaussian rest psi0}
\end{align}
with rest energy  centered at $p^0=m>0$ with spread $\Delta_0>0$. This state is defined such that the clock's 3-momentum distribution is centered at ${\bf p} =0$ and, thus, it is (on average) at rest in our global inertial coordinates $x^\mu=(t,x,y,z)$ with with variance $\Delta^2>0$. Here, $N$ is a normalization constant and $\chi$ is the indicator function of $\sigma(H_C)$,

\begin{align}
    \chi(E) \equiv \begin{cases}
        1, \, E \in [-\Omega,\Omega] \\
        0, \, E \not\in [-\Omega,\Omega]
    \end{cases}\, .
\end{align}
Evolving the the initial state \eqref{gaussian rest psi0} via Eqs.~\eqref{evolved state} and~\eqref{evolved state 2}, we obtain its ``worldfunction" 

\begin{align}
    \psi(x^\mu;\tau) &= \frac{N}{4\pi^2} \int d^3 \mathbf{p} \, dp_0 \, e^{-\frac{(p_0-m)^2}{2\Delta_0^2}}e^{-\frac{\mathbf{p}^2}{2\Delta^2}} \nonumber \\ 
    & \times e^{-i p_\mu x^\mu} e^{-i(m-\sqrt{p^2})\tau} \chi(\sqrt{p^2}-m).
\end{align}
In order to get a better understanding of the physics codified in the above worldfunction, let us analyze it in a weakly relativistic regime. In such a regime, the amplitudes are relevant only in the regions $\mathbf{p}^2 \ll m^2$ and $|q_0| \equiv |p_0 - m| \ll m$. Hence, in therms of our parameters, this implies that $\Delta_0, \Delta \ll m$. Furthermore, if we impose that the Gaussians are sufficiently tight in momentum space with respect to the clock's spectrum, i.e., $\Delta_0, \Delta \ll \Omega$, we can ignore the indicator function $\chi$ and perform a simple Fourier Transform in $\mathbb{R}^4$ obtaining

\begin{align}
    \psi(x^\mu;\tau) \propto e^{imt} e^{-\frac{\Delta_0^2}{2}(t-\tau)^2} \bigl( \Delta^{\!\prime\, 2}(\tau) \bigl)^{3/2} e^{-\frac{1}{2}\Delta^{\!\prime\, 2}(\tau)\mathbf{x}^2}, \label{psitau de x}
\end{align}
where we have defined

\begin{align}
    \Delta^{\!\prime\, 2}(\tau) \equiv \frac{\Delta^2}{1-i\Delta^2\tau/m}.
\end{align}

Thus, unlike the case of plane-waves, the solution described Eq.~\eqref{psitau de x} is spacetime-localized, with a worldfunction that peaks at the event $x^\mu(\tau)=(t=\tau; \mathbf{x}=\mathbf{0})$ for each proper time $\tau$. Hence, the worlddunction is peaked around classical worldline of an inertial particle at rest at $\mathbf{x}=\mathbf{0}$. Its spatial dependence displays the usual behavior of a free particle in non-relativistic quantum mechanics (NRQM), namely, a 3-dimensional Gaussian diffusing in (proper) time. The temporal component is just a Gaussian translating uniformly with $\tau$, keeping the worldfunction (temporally) localized around $t=\tau$.

\subsection{Multiple clocks: global constraints and commensurability of time evolution}

Let us now consider the case of multiple non-interacting clocks. Recall that in the classical theory, each relativistic particle has one independent Hamiltonian constraint in phase space, given in Eq.~\eqref{independent local constraints}. Similarly, for the quantized theory, the physical Hilbert space for $N$ clocks will be comprised of states $\Pket$ simultaneously obeying the $N$ Hamiltonian constraints
\begin{align}
    H_I\Pket=0, \qquad I=1,\dots N.
\end{align}
This can always be satisfied for non-interacting clocks, as all constraints commute, $[H_I,H_J]=0$. If the set of constraints does not commute, the formalism presented here (and in other works~\cite{flaminiaSTQRFs,smithQC1}) is not sufficient, and further work would be needed.

As a result, not only the kinematical Hilbert space will be a tensor product of each of the clock spaces,
\begin{align}
    \mathcal{H}_{kin} = \bigotimes_{I=1}^N \mathcal{H}_{kin}^{(I)}, 
\end{align}
but also the physical Hilbert space will share this simple decomposition:
\begin{equation}
    \mathcal{H}_{phys} = \bigotimes_{I=1}^N \mathcal{H}_{phys}^{(I)}. 
\end{equation}
One important consequence of this factorization of the physical Hilbert space is that one can immediately obtain a basis of global solutions in $\mathcal{H}_{phys}$ from products of the local solutions of the individual constraints $\dket{\Psi^j_I} \in \mathcal{H}^{(I)}_{phys}$, $j\in \mathbb{N}.$ However, similarly to what happened in case of a single clock, we must be wary that this Hilbert space still bears many unphysical solutions (from the perspective of the phenomenology of our usual, well-tested quantum theories such as NRQM and QFT). 

Note that, for separable solutions, one does not have any correlation between the clocks' proper times $(\tau_I,\tau_J)_{I \neq J}$, regardless of the clocks' individual states of motion $\rho_I, \rho_j$. For instance, we could have two clocks prepared in identical semiclassical states of the form \eqref{psitau de x}, so that both are centered on identical inertial worldlines $x^\mu_I(\tau_I) = (t_I=\tau_I;\,\mathbf{x}_I=\mathbf{0})$ (with their worldfunctions overlapping in their entire histories), and still their proper times would be completely uncorrelated with each other. The global joint probabilities of measuring each clock at arbitrary intervals $I_I$ breaks as a product

\begin{widetext}

\begin{align}
    P_\rho(\tau_1\!\in\! I_1, \tau_2\!\in\! I_2) &= \nu_1 \nu_2 \int_{I_1} d\tau_1 \int_{I_2} d\tau_2 \Tr\{\rho \,\ketbra{\tau_1,\tau_2}{\tau_1,\tau_2} \} \nonumber \\
    &= \biggl( \nu_1\!\int_{I_1} d\tau_1 \Tr\{ \dket{\Psi_1}\dbra{\Psi_1}\ketbra{\tau_1}{\tau_1} \} \biggl) \biggl( \nu_2\!\int_{I_2} d\tau_2 \Tr\{ \dket{\Psi_2}\dbra{\Psi_2}\ketbra{\tau_2}{\tau_2} \} \biggl) \nonumber \\[4pt]
    &= P_{\rho_1}(\tau_1 \in I_1) \,P_{\rho_2}(\tau_2 \in I_2),
\end{align}

\end{widetext}
where $\rho\equiv \rho_1\otimes \rho_2.$

Thus, to ensure that our physical states have a commensurable notion of time evolution, we adopt the further global
constraint that the external temporal DoFs $t_I$ are strongly correlated with one another. In particular, in order to make them completely correlated, we follow~\cite{smithQC1} and apply the improper projector 
\begin{align}
    P_{\mathbf{\Sigma}} \equiv \int dt \ketbra{t}{t},
\end{align}
with $\ket{t}$  defined as
\begin{align}\label{Psigma}
    \ket{t} &\equiv \bigotimes_{I=1}^N \ket{t_I=t} \nonumber \\
    &= \int \biggl[\prod_{I=1}^{N-1} dt_I \delta(t_I-t_{I+1}) \biggl]\delta(t-t_1)\ket{t_1,\dots t_N},
\end{align}
in the clocks states.  Heuristically, this can be thought of as a Hilbert space analog of choosing a spacetime foliation $\mathbf{\Sigma}\equiv \{\Sigma_t\}_{t\in \mathbb{R}}$ (thus the notation $P_{\mathbf{\Sigma}}$)\footnote{We stress, however, that the analogy with choosing a spacetime foliation is not perfect.  Applying $P_{\mathbf{\Sigma}}$ is an active process in $\mathcal{H}_{phys}$ which allows us to obtain temporally correlated clock states from separable ones.}, and states that have the form $\dket{\Psi_{\mathbf{\Sigma}}}=P_{\mathbf{\Sigma}}\Pket$ can be more easily interpreted in terms of states that are prepared at a Cauchy surface $\Sigma_{t_i}$ and measured at a Cauchy surface $\Sigma_{t_f}$. Of course, one needs not to restrict oneself to perfectly correlated time-variables. we
simply want to make sure that there are nontrivial correlations between the clocks and that there
is a somewhat common notion for time-evolution. For this more general case, one can substitute the delta distribution in Eq.~(\ref{Psigma}) for some fuzzier distribution, $\Delta(t)$, peaked at $t=0$.

For our globally constrained physical states, we define a global notion of time evolution ``in the perspective'' of a given clock as follows: we start with the symmetric history state 

\begin{align}
    \dket{\Psi_{\mathbf{\Sigma}}} = \int \Bigl[\bigotimes_{J=1}^N d\tau_J\ket{\tau_J} \Bigl] \ket{\psi_\Sigma(\tau_1,\dots \tau_N)}, 
\end{align}
where
\begin{align}
    \ket{\psi_{\mathbf{\Sigma}}(\tau_1,\dots \tau_N)} = P_{\mathbf{\Sigma}} \bigotimes_{J=1}^N \ket{\psi_J(\tau_J)}.
\end{align}
Then, we pick a fiducial clock $I$, factor-out its internal DoF  to serve as a temporal reference frame and write a conditional state for all remaining DoFs (internal and external DoFs of the remaining clocks, $J\!\neq\! I$, plus the external DoFs of clock $I$ itself) as

\begin{align}
     \dket{\Psi_{\mathbf{\Sigma}}} 
     &=\int d\tau_I\ket{\tau_I} \ket{\Psi'_{\mathbf{\Sigma}}(\tau_I)}. \label{global TE}
\end{align}
with
\begin{align}
     \ket{\Psi'_{\mathbf{\Sigma}}(\tau_I)} &=  \int \Bigl[\bigotimes_{J\neq I} d\tau_J\ket{\tau_J} \Bigl]
     \ket{\psi_{\mathbf{\Sigma}}(\tau_1,\dots \tau_N)}.
     \label{global TEb}
\end{align}
Here, we leave ${\mathbf{\Sigma}}$ explicit in our conditional state $\ket{\Psi'_\Sigma(\tau_I)}$ to stress that it is through this foliation choice that we are able to extrapolate from a local notion of time evolution involving only clock $I$ to a global one involving all clocks. We note, however, that this this definition of simultaneity (i.e., the choice of $P_{\mathbf{\Sigma}}$) still leaves room for quantum fluctuations due to internal fluctuations of the clocks.

Now, the state in Eq.~(\ref{global TE}) allows us to define a notion of ``quantum time-dilation'' by comparing the proper-time variables $\tau_J$ and $\tau_I$ of two distinct clocks
$I\!\neq\! J$ by means of the conditional density operator 

\begin{align}
    \rho_{\tau_J|\tau_I} \equiv \Tr_R \bigl\{ \ketbra{\Psi'_\Sigma(\tau_I)}{\Psi'_\Sigma(\tau_I)} \bigl\},
\end{align}
where this partial trace $\Tr_R$ is taken over all the DoFs of our conditional state~(\ref{global TE}), except for the  internal DoF of clock $J$. In the next section, we shall compute this conditional density operator for some different paradigmatic applications.

\section{Applications}\label{sect:Applications}

Now that we have laid down our framework, let us apply it to some simple systems of interest.

\subsection{Inertial clocks at rest relative to each other}
As a first example, let us consider two identical free clocks prepared in the same Gaussian state, i.e., at rest with respect to each other. This yields the simplest analytical expressions while still displaying nontrivial quantum time dilation.  In this case, the worldfunction of each of the clocks is given by Eq.~\eqref{psitau de x}, hence

\begin{align}
    \psi_{\tau_I}(x^\mu_I;\tau_I)\!=\!N_J e^{imt_I} e^{-\frac{\Delta_0^2}{2}(t_I-\tau_I)^2} \bigl( \Delta^{\!\prime\, 2}(\tau_I) \bigl)^\frac{3}{2} e^{-\frac{\Delta^{\!\prime\, 2}(\tau_I)}{2}\mathbf{x}_I^2} \label{psitau_I de x}, \quad 
\end{align}
with $N_J$ being a normalization constant and $I=1,2.$ Then, by using clock 1 as the fiducial clock, our conditional state~\eqref{global TE} reads

\begin{align}
    &\!\!\ket{\Psi'_\Sigma(\tau_1)} \!=\!\!\int \!d\tau_2 \ket{\tau_2} \!\int \!d^3\mathbf{x}_1 d^3\mathbf{x}_2 \, dt \ket{t_1\!=\!t,t_2\!=\!t}\ket{\mathbf{x}_1}\ket{\mathbf{x}_2} \nonumber \\
     &\times \mathcal{N} e^{2imt}\! \prod_{I=1,2} e^{-\frac{\Delta_0^2}{2}(t-\tau_I)^2} \bigl( \Delta^{\!\prime\, 2}(\tau_I) \bigl)^{\!\frac{3}{2}} e^{-\frac{1}{2}\Delta^{\!\prime\, 2}(\tau_I)\mathbf{x}_I^2},
\end{align}
where $\mathcal{N}\equiv N_1N_2$. Finally, taking the partial trace in all external variables ($\mathbf{x}_1,\mathbf{x}_2,t$), we arrive at the conditional density operator

\begin{align}
    \rho_{\tau_2|\tau_1}\! \!\propto \!\!\int \!\!d\tau_2' d\tau_2'' \ket{\tau_2'}\!\bra{\tau_2''} e^{ -\frac{\Delta_0^2}{8} \bigl[ (\tau_2'-\tau_2'')^2 + 2\bigl( (\Delta\tau_2')^2 + (\Delta\tau_2'')^2 \bigl) \bigl] }, \label{rest QTD}
\end{align}
where $\Delta\tau_2 \equiv \tau_2-\tau_1$. 

The physical interpretation of Eq.~\eqref{rest QTD} is straightforward. First, if we take the diagonal terms, $\tau_2'=\tau_2''$, we see that the probability density for measuring a proper time time $\tau_2$ in clock 2, given that clock 1 is in state $\ket{\tau_1}$, is simply a Gaussian peaking at $\tau_2=\tau_1$

\begin{align}
    P(\tau_2|\tau_1) \propto e^{ -\frac{\Delta_0^2}{2}(\tau_2-\tau_1)^2}.
\end{align}
\begin{figure}[H]
    \includegraphics[width=0.8\linewidth]{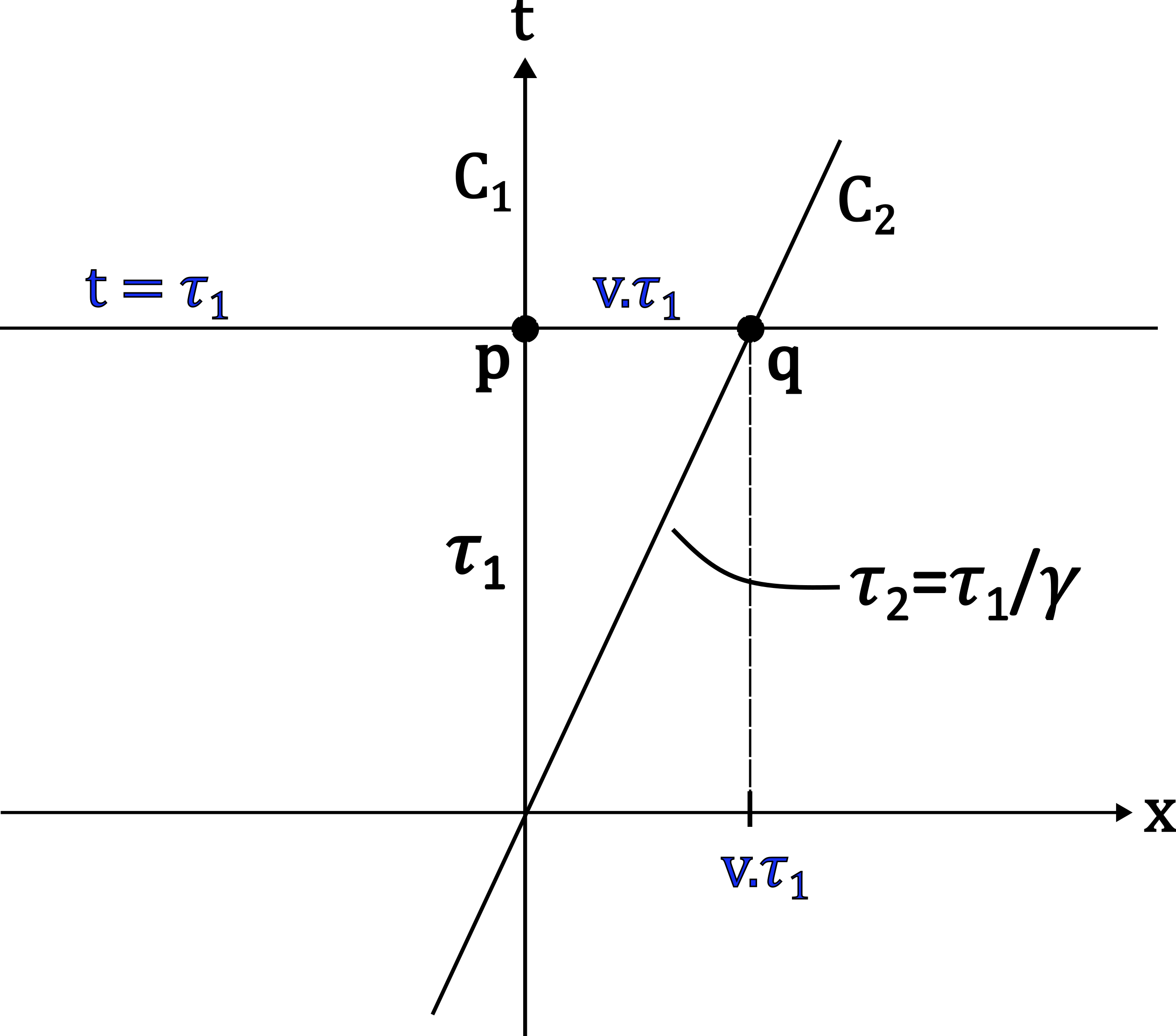}
    \caption{Classical time-dilation between clocks with relative 3-velocity $v$, evaluated for spacelike separated events $p$ and $q$, with respect to the simultaneity surface $\{t=\tau_1\}$ orthogonal to the 4-velocity of clock 1. Clocks 1 and 2 read proper times $\tau_1$ and $\tau_2=\tau_1/\gamma$ at events $p$ and $q$, respectively. Our conditional density operator $\rho_{\tau_2|\tau_1}$ peaks exactly at the expected classical values, but also presents quantum coherences and fluctuations.}
    \label{inertialTD}
\end{figure}
Furthermore, there are nontrivial off-diagonal terms ($\tau_2'\!\neq\!\tau_2''$) showing that the temporal superposition of clock 2 relative to clock 1 is coherent--justifying the term quantum time dilation. In this simple example, the coherences also have a Gaussian behavior, decaying as $e^{ -\frac{\Delta_0^2}{8} (\tau_2'-\tau_2'')^2}$ (relative to the populations) as we move away from the diagonal.


\subsection{Inertial clocks with finite relative velocity}
\label{sec:2InertialClocks}

The next example we consider is two inertial clocks with a finite relative 3-velocity $\mathbf{v}$ between them. This is simple to obtain from our previous example by applying a Lorentz boost to, e.g., the state of clock 2. Hence, Clock 1 ( which we will once again take as a fiducial clock) is prepared in the state 

\begin{align}
  \!\!  \psi_{1}(x^\mu_1;\tau_1) =N_1 e^{imt_1} e^{-\frac{\Delta_0^2}{2}(t_1-\tau_1)^2} \bigl( \Delta^{\!\prime\, 2}(\tau_1) \bigl)^\frac{3}{2} e^{-\frac{\Delta^{\!\prime\, 2}(\tau_1)}{2}\mathbf{x}_1^2} \label{psitau_1 de x},
\end{align}
whereas Clock 2 will be prepared in a boosted state $\ket{\psi'}=U(\mathbf{v})\ket{\psi}$, for which we shall pick $\mathbf{v}$ along the $x$-axis, $\mathbf{v}=v\hat{\mathbf{x}}$. As a result, 

\begin{align}
    &\psi'(x_2;\tau_2) =N_2 e^{im\gamma(t_2-vx_2)} e^{-\frac{\Delta_0^2}{2}\bigl[\gamma(t_2-vx_2)-\tau_2\bigl]^2} \nonumber  \\
    &\times   \bigl( \Delta^{\!\prime\, 2}(\tau_2) \bigl)^{3/2} e^{-\frac{1}{2}\Delta^{\!\prime\, 2}(\tau_2)\mathbf{x}_{2\perp}^2}e^{-\frac{1}{2}\Delta^{\!\prime\, 2}(\tau_2)\gamma^2(x_2-vt_2) },  \label{psitau' de x}
\end{align}
where $\mathbf{x}_{2\perp}$ denotes the transverse components of $\mathbf{x}_2$ in the $yz$ plane and $\gamma$ is the usual Lorentz factor $\gamma=(1-v^2)^{-\frac{1}{2}}$.

We now choose a sharp global constraint (``foliation'') $P_{\mathbf{\Sigma}}$ privileging the time coordinate $t$ for which clock 1 is (on average) at rest

\begin{align}
    P_{\mathbf{\Sigma}}
    = \int dt_1 dt_2 \ketbra{t_1,t_2}{t_1,t_2} \delta(t_1-t_2).
\end{align}
As a result the conditional state reads

\begin{widetext}

\begin{align}
    &\ket{\Psi'(\tau_1)} =\mathcal{N} \int d\tau_2 \ket{\tau_2} \int (d^3\mathbf{x}_1) (d^2\mathbf{x}_{2\perp}) \, dxdt \ket{\mathbf{x}_1}\ket{x_2\!=\!x,\mathbf{x}_{2\perp}} \ket{t_1\!=\!t,t_2\!=\!t} \nonumber \\
     \times& e^{imt} e^{im\gamma(t-vx)}  \bigl( \Delta^{\!\prime\, 2}(\tau_1) \bigl)^{3/2} e^{-\frac{1}{2}\Delta^{\!\prime\, 2}(\tau_1)\mathbf{x}_1^2} \bigl( \Delta^{\!\prime\, 2}(\tau_2) \bigl)^{3/2} e^{-\frac{1}{2}\Delta^{\!\prime\, 2}(\tau_2)\mathbf{x}_{2\perp}^2}e^{-\frac{1}{2}\Delta^{\!\prime\, 2}(\tau_2)\gamma^2(x-vt)^2}   
     e^{-\frac{\Delta_0^2}{2}(t-\tau_1)^2} e^{-\frac{\Delta_0^2}{2}\bigl[\gamma(t-vx) -\tau_2 \bigl]^2}, \label{cond state inertial w rel motion}
\end{align}
yielding the conditional density matrix 

\begin{align}
    \rho_{\tau_2|\tau_1} \propto \int \!d\tau_2' d\tau_2'' \ket{\tau_2'}\!\bra{\tau_2''}
    \exp\biggl( -\xi^2\bigl[ \tilde{\Delta}^2(\tau_2'-\tau_2'')^2 + 2\gamma^2\Delta^2\bigl( (\Delta\tau_2^{\prime})^2 + (\Delta\tau_2^{\prime\prime})^2   \bigl)  \bigl] \biggl). \label{rho2|1 w rel motion}
\end{align}
\end{widetext}
Here we have defined
\begin{eqnarray}
\Delta\tau_2 &\equiv& \tau_2- \tau_1/\gamma \\
    \Delta^2 &\equiv& \frac{1}{2}\bigl( \Delta^{\!\prime\,2}(\tau_2') + \Delta^{\!\prime\,2}(\tau_2'') \bigl), \\
     \tilde{\Delta}^2 &\equiv& \Delta^2 + (\gamma v \Delta_0)^2, \\
    \xi^2 &\equiv& \frac{\Delta_0^2}{(1+\gamma^2)\Delta^2 + \gamma^2v^2\Delta_0^2} = \frac{\Delta_0^2}{ \tilde{\Delta}^2 + \gamma^2\Delta^2} .
\end{eqnarray}

We note that the conditional probabilities $P(\tau_2|\tau_1)$ arising from the diagonal terms of $\rho_{\tau_2|\tau_1}$ in Eq.~\eqref{rho2|1 w rel motion} peak around the classical value of time dilation $\tau_2=\tau_1/\gamma$ with respect to the foliation $\Sigma_t$ orthogonal to the 4-velocity of clock 1 (see Fig.~\ref{inertialTD}). As in the previous example, the density matrix~\eqref{rho2|1 w rel motion} exhibit Gaussian coherences codified in its off-diagonal elements.

\subsection{Charged clocks in a uniform magnetic field}

Let us now consider the case where our clocks are accelerated via a coupling with a uniform magnetic field. For this purpose, we shall assume that they are charged with charge $q<0$. This case is interesting both because it is very simple to treat analytically and because it should be fairly simple to implement experimentally. Furthermore, charged particles in such a field configuration remain in a bounded region of space, following circular orbits (assuming they are prepared with zero momentum parallel to the field). This enables one to compare the time dilation between a pair of clocks locally in a compact laboratory rather than in spatially separated events (since its worldlines can be arranged to intersect periodically), foregoing the need for an arbitrary choice of simultaneity surfaces .

We shall describe our magnetic field as being static and uniform in our set of global inertial coordinates $\{t,x,y,z\}$, pointing along the $z$-axis, $\mathbf{B}=B_0\hat{\mathbf{z}}$, with $\hat{\mathbf{z}}$ being the unit field along the $z$ direction. We then choose a convenient gauge for the vector potential $A^\mu(x)$ such that its components, in the coordinates $\{t,x,y,z\}$,  read

\begin{align}
    \begin{cases}
        A^0 = 0, \\[4pt]
        \mathbf{A} = -\frac{1}{2}\mathbf{B}\times\mathbf{x}
    \end{cases} \label{UMF gauge}
\end{align}
In this gauge, the quadratic Hamiltonian constraint of each clock is given by

\begin{align}
    H &\propto M^2 - [P-qA(x)]^2 \nonumber \\[3pt]
    &= M^2 - (P^0)^2 +(P_{z})^2 + [\mathbf{P}_{\perp} - q\mathbf{A}(\mathbf{x})]^2  \label{H  UMF}
\end{align}

It turns out that, in this case, it is more convenient to solve the quadratic constraint $H$, rather than the linearized $H_{-}$ (as we mentioned in Sec.~ \ref{sec:Quantization}, both constraints are equivalent if $\sigma(M) \subset \mathbb{R}^+$ and $[H_{-},H_{+}]=0$ in $\mathcal{H}_{kin}$) since the transverse DoFs naturally yield a 2D harmonic oscillator. We refer the reader to App.~\ref{ap:UMF} for a detailed construction of the constraint solutions (and of the individual coherent states built from them) that will be used in this section.

\begin{figure}[H]
    \centering \hspace{0.1\linewidth}
    \includegraphics[width=0.4\linewidth]{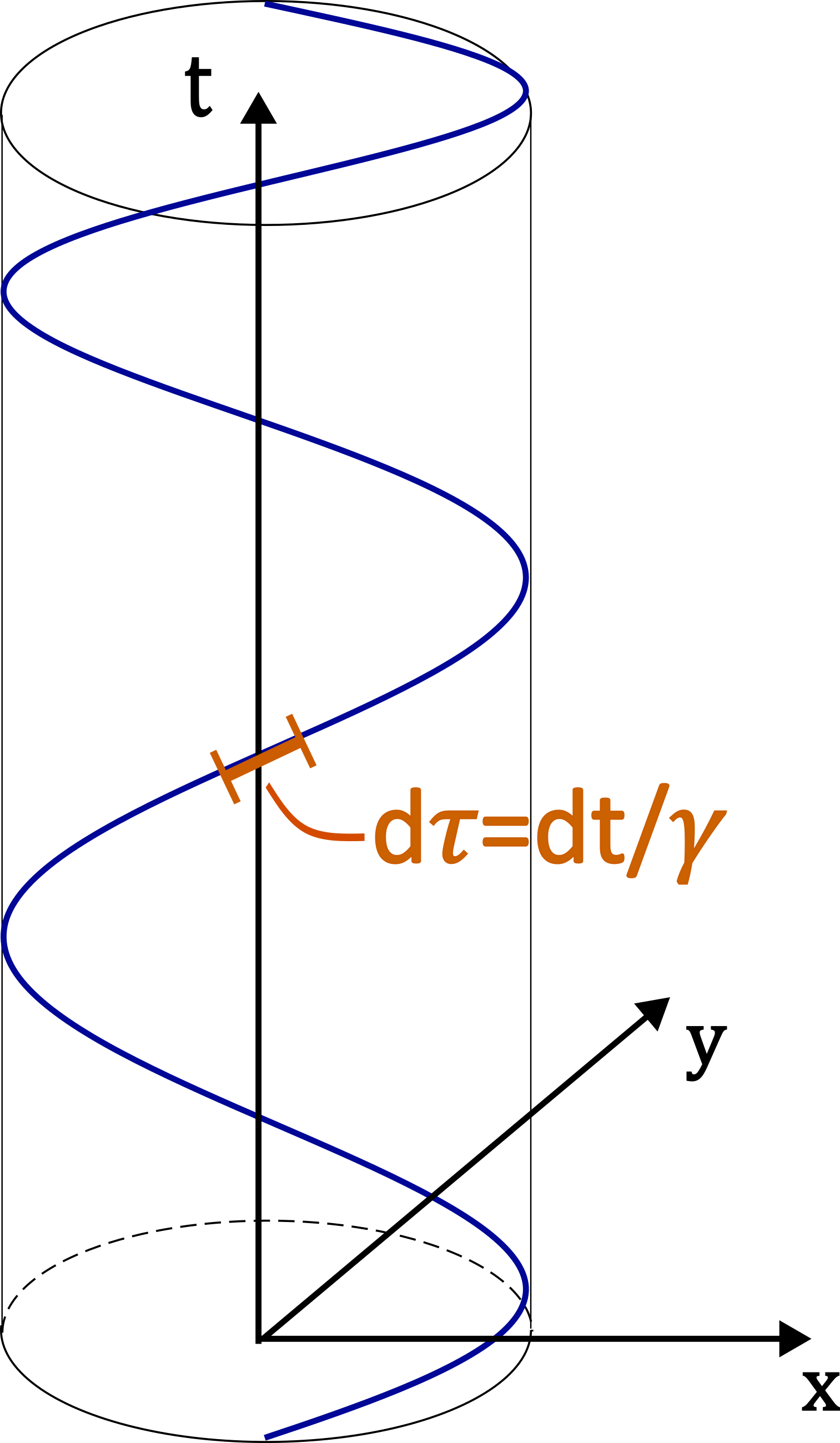}
    \caption{Classical worldline of a charged particle in a uniform magnetic field. The particle's proper time $\tau$ as a function of the coordinate time $t$ is simply $\tau=t/\gamma$, where $\gamma$ can be expressed as a function of the energy, $\gamma=E/m$. }
    \label{mag-TD}
\end{figure}

Analogously to inertial clocks, it is convenient to choose quasi-classical states combining Gaussians in the $t$ and $z$ subspaces with coherent states $\ket{\alpha_r, \alpha_l}$ [see Eq.~(\ref{coherent r l})] for the transverse direction. However, as explained in App.~\ref{ap:UMF}, since the left-polarized quanta do not contribute to the total energy when $q<0$, we simply take $\alpha_l=0$. As a result [see derivation leading to Eq.~\eqref{app NR UMF psi(tau)}], the time evolution of the clock with respect to the clock´s proper time $\tau$ is given by 

\begin{align}
   \ket{\psi(\tau)}&= N \biggl[ \int\! dt dz \ket{t,z} e^{iEt} e^{-\frac{\Delta_0'^2(\tau)}{2}\!\bigl[t - \bar{\gamma}\tau\bigl]^2}\! e^{-\frac{\Delta_z'^2(\tau)}{2}z^2} \biggl] \nonumber \\
   & \otimes \Ket{ \alpha_r e^{-i\omega_c\tau}, \alpha_l\!=\!0  }e^{i|\alpha_r|^2\omega_c\tau},
\end{align}
where we have defined an effective $\gamma$-factor $\bar{\gamma} \equiv E(\alpha_r)/m$, with 
\begin{equation}
 E(\alpha_r) \equiv \sqrt{m^2+2m\omega_c\left(|\alpha_r|^2+ 1/2\right)^2}    
\end{equation}
and $\omega_c\equiv |q|B_0/m$, which we associate to an effective velocity 
\begin{equation}
    \bar{v}\equiv\sqrt{E^2(\alpha_r)-m^2}/E(\alpha_r).
\end{equation} 

\vspace{20pt}
 \begin{figure}[H]
    \centering \vspace{12pt} 
    \includegraphics[width=0.9\linewidth]{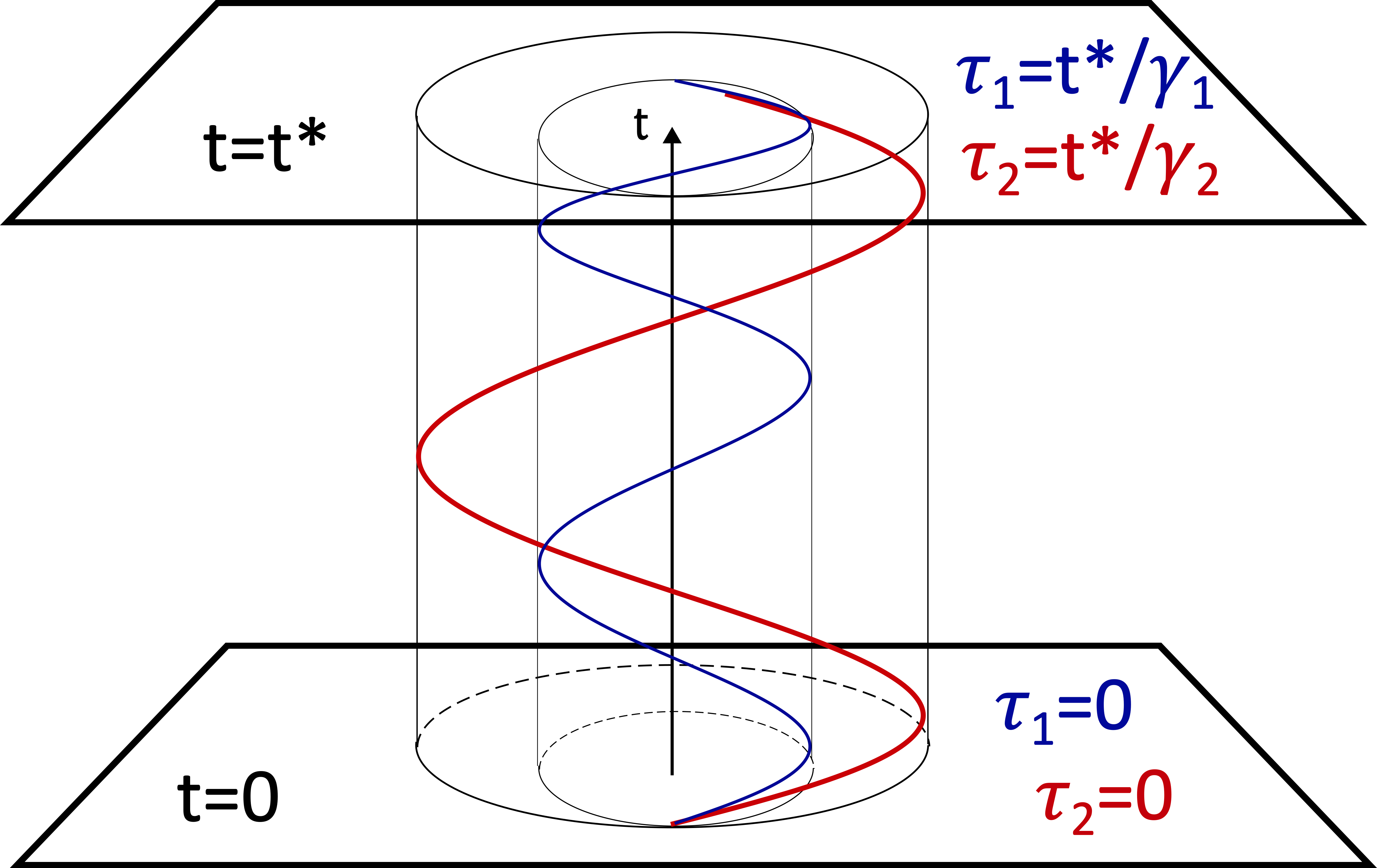}
    \caption{Classical worldlines of two charged particles in a uniform magnetic field. If both particles carry a clock synchronized at $\tau_1=\tau_2=0$ at $t=0$, they will read $\tau_I=t^*/\gamma_I$ at $t=t^*$. Furthermore, if the particles' motions are arranged properly, their worldlines will intersect periodically at certain times, so that one can compare the proper time difference locally, independently of the choice of foliation.}
    \label{mag-TD2}
\end{figure}
Here, 
\begin{align}
    \Delta_0'^2(\tau)&=\frac{\Delta_0^2}{1-i\Delta_0^2\,\bar{v}^2\bar{\gamma}^2\tau/m}, \\
    \Delta_z'^2(\tau)&= \frac{\Delta_z^2}{1-i\Delta_z^2\tau/m},
\end{align}
with $\Delta_0$ $\Delta_z>0$ being the initial temporal and z-direction Gaussian spreads. These states are very straightforward to interpret in terms of a quasi-classical motion, see Figure~\ref{mag-TD}). In particular, they display the expected average time dilation with respect to the inertial time coordinate for which the field is static and uniform.

Now, it is straightforward to evaluate the conditional time evolution of two charged clocks, each prepared in its own coherent state $\ket{\alpha^I_{r}}$, $I=1,2$.  If we take clock 1 as our fiducial clock and choose a foliation privileging the inertial time coordinate $t$, by an analogous computation as the one done in Sec.~\ref{sec:2InertialClocks} we obtain the conditional state

\begin{widetext}
    
\begin{align}
    \ket{\Psi(\tau_1)} =\mathcal{N} \int d\tau_2 \ket{\tau_2}
    \bigotimes_{I=1,2} \Biggl[ \biggl( &\int dt \ket{t_I\!=\!t} e^{iE_It} e^{-\frac{\Delta_0'^2(\tau_I)}{2}\bigl[t - \bar{\gamma}_I\tau_I \bigl]^2} \biggl) \nonumber \\
     &\biggl( \int dz_I \ket{z_I} e^{-\frac{\Delta_z'^2(\tau_I)}{2}z_I^2 } \biggl) 
     \Ket{\alpha_{r}^Ie^{i{\omega_c\tau_I}}, \alpha_{l}^I=0 } e^{i{\omega_c\tau_I}|\alpha_{r}^I|^2} \Biggl],
\end{align}
from which we can compute the conditional density matrix
\begin{align}
    \rho_{\tau_2|\tau_1} \propto \int d\tau_2' d\tau_2'' \ket{\tau_2'}\!\bra{\tau_2''} &e^{-\frac{\Delta_2^2}{2\Tilde{\Delta}_0^2} \Bigl[ \Delta_2^2\bar{\gamma}_2^2(\tau_2'-\tau_2'')^2 + 2 \Delta_1^2 \bigl( (\bar{\gamma}_2\tau_2'-\bar{\gamma}_1\tau_1)^2 + (\bar{\gamma}_2\tau_2''- \bar{\gamma}_1\tau_1)^2 \bigl) \Bigl]   } \nonumber \\
    &\times e^{i|\alpha^2_{r}|^2\bigl[\omega_c(\tau_2'-\tau_2'')-\sin\bigl(\omega_c(\tau_2'-\tau_2'')\bigl)\bigl]}e^{|\alpha^2_{r}|^2\bigl[-1+ \cos\bigl(\omega_c(\tau_2'-\tau_2'')\bigl) \bigl]}  \label{rho2|1 mag3}
\end{align}
\end{widetext}
where we have defined $\Delta_1 \equiv \Delta'_0(\tau_1)$, $\Delta'_2 \equiv \Delta'_0(\tau'_2)$ and $\Delta''_2 \equiv \Delta'_0(\tau''_2)$, as well as an average spread $\Tilde{\Delta}_0 = 2\Delta_1 + \Delta'_2 + \Delta''_2$, and we have approximated $\Delta_2' \simeq \Delta_2'' \equiv \Delta_2$.

Once again, we find that the conditional probabilities given by the diagonal values of $\rho_{\tau_2|\tau_1}$ ($\tau_2'=\tau_2''$) peak exactly at the classical time- dilation values $\bar{\gamma}_2\tau_2=\bar{\gamma}_1\tau_1$ (see Fig.~\ref{mag-TD2}) and decay with a Gaussian profile. {Its coherences, in turn, display an additional feature: besides the Gaussian factor, they also present periodic modulation of their amplitude and phase as a function of the time difference $\tau_2'-\tau_2''$.}

Finally, we note that, if the initial states of our clocks is prepared appropriately, their worldfunctions can be made to overlap periodically. Then, even though our calculations were made in a particular inertial foliation, this choice becomes irrelevant if the time dilation is evaluated locally in the overlap regions\footnote{As we have finitely spread center of masses, rather than idealized classical point particles, this local comparison ought to be considered as occurring in a sufficiently compact spatial region, for which any foliation dependence is vanishingly small.}.


\section{Discussion} \label{sec:conclusions}

In this paper, we have proposed a new spacetime-covariant formalism  to describe quantum clocks in first quantization. This can be done whenever de (quadratic) clock Hamiltonian can be decomposed in terms of positive- and negative-mass sectors. When such a decomposition is attainable, we can compute the evolution of our systems directly with respect to the clock's proper time while maintaining explicit covariance. We were able to show that, contrary to previous claims in the literature, it is possible to recover an (emergent) unitary evolution for the CM DoF of an accelerated clock (for instance, by accelerating a (charged) clock with an arbitrary electromagnetic field). 

We have also extended the formalism to cover the case of multiple clocks. To this end, we impose extra constraints on the physical states of our theory to obtain a commensurable notion of time evolution. By choosing a fiducial reference clock $I$, one is able to obtain the (unitary) conditional time evolution of all remaining degrees of freedom in terms of the its proper time $\tau_I$ and then, from such a conditional state, we can obtain the conditional density operator $\rho_{\tau_I|\tau_J}$ describing time measurements made by clocks $I$ and $J\neq I$, which allows us to quantify time dilation between them.

We have ended the paper with three paradigmatic applications of our formalism, namely: {\bf (A)} two inertial clocks at rest relative to each other,  {\bf (B)} two inertial clocks with relative velocity, and {\bf (C)} two accelerated (charged) clocks in a uniform magnetic field. In all three cases, one obtains the usual classical results through the peak values of the diagonal terms of  $\rho_{\tau_I|\tau_J}$ (i.e.,  $\tau_1=\tau_2$ for case (A) and $\tau_1=\gamma \tau_2$ and $\overline\gamma_1\tau_1=\overline\gamma_2\tau_2$ for the cases (B) and (C), respectively, with $\gamma, \overline\gamma_1,$ and $\overline\gamma_2$ being the suitable Lorentz factor in each case). In all cases, however, there are quantum fluctuations codified in the off-diagonal terms of the density matrix.

\acknowledgments
EO research was fully funded by  S\~{a}o Paulo Research Foundation (FAPESP) under the Grants \#2022/00034-1 and \#2023/05830-3.

\appendix

\section{Charged particle in a Uniform Magnetic Field} \label{ap:UMF}

Classically, a particle with charge $q$ in a uniform magnetic field $\mathbf{B} = B_0 \hat{\mathbf{z}}$ will have a transverse motion in the $xy$-plane corresponding to circular orbits with cyclotron frequency $\omega_c\equiv \frac{|q|B_0}{m}$. For $q<0$ these orbits will be counterclockwise, whereas for $q>0$ they will be clockwise. In the quantized theory, it is convenient to work with global inertial coordinates and the gauge \eqref{UMF gauge}, for which the vector potential `circulates' around the z-axis. With this choice, the magnetic contribution to the Hamiltonian is isolated in a transverse term\footnote{More precisely, we define the operator $C_\perp \equiv \gamma(s)^{-1}MH_\perp$ for convenience, so that the transverse external DoFs do not mix with the internal ones. The separable quadratic constraint $C\equiv \gamma(s)^{-1}MH = M^2-(P^\mu-qA^\mu)^2$ is equivalent to $H$ and $H^-$ for a strictly positive mass spectrum $\sigma(M)>0$.}
\begin{equation}
H_\perp \propto C_\perp = (\mathbf{P}_\perp-q\mathbf{A}(\mathbf{x}))^2 
\end{equation}
and the z-component of the canonical angular momentum $\mathbf{L}=\mathbf{x}\times \mathbf{P}$ is a constant of motion (i.e., it commutes with every term of the Hamiltonian constraint). In this gauge, $C_\perp$ takes the form of a 2D harmonic oscillator

\begin{align}
    C_\perp = \mathbf{P}_\perp^2 + m^2\biggl( \frac{\omega_c}{2} \biggl)^2\mathbf{x}_\perp^2 - \operatorname{sgn}(q)\, m\omega_cL_z,
\end{align}
where $\operatorname{sgn}(x)= \frac{x}{|x|}$ denotes the sign function. This oscillator can be diagonalized in terms of right and left circularly polarized quanta, with number operators $N_r = a^\dagger_r a_r$ and $N_l = a^\dagger_l a_l$, respectively. They relate to the $x$ and $y$ linearly polarized quanta by
\begin{align}
    a_r = \frac{1}{\sqrt{2}}(a_x-ia_y) ; \qquad
    a_l = \frac{1}{\sqrt{2}}(a_x+ia_y),
\end{align}
with $a_x, a_y$ being the usual annihilation operators. They form a complete set of commuting operators in the transverse sub-space. Its eingenstates $\ket{n_r,n_l}$, $n_r,n_l \in \mathbb{N}$, form a basis for such a sub-space with 
\begin{align}
    \hspace{-3pt}
    C_\perp \ket{n_r,n_l} = \mu_c^2 \!\times\!
    \begin{cases}
    \bigl(n_r+ \frac{1}{2}\bigl) \ket{n_r,n_l}, \text{ for } q<0, \\[2pt]
    \bigl(n_l+ \frac{1}{2}\bigl) \ket{n_r,n_l}, \text{ for } q>0,
    \end{cases}
\end{align}
where $\mu_c^2 \equiv 2m\omega_c=2|q|B_0$,  and well-defined axial angular momentum
\begin{align}
    L_z\ket{n_r,n_l} = (n_r-n_l)\ket{n_r,n_l}.
\end{align}
 Note that the quanta that contribute to the $C_\perp $ (and, hence, to $H_\perp$) eigenvalues are precisely those corresponding to the classical circular orbits for either positive or negative charges, and all energy levels are infinitely degenerate, as they do not depend on the complementary polarized quanta. For definiteness, we shall take $q<0$ from here on, such that $C_\perp=(N_r+1/2)\omega_c$.

From the number states $\ket{n_r,n_l}$, we can build coherent (quasi-classical) states $\ket{\alpha_r,\alpha_l}$ defined as

\begin{align}\label{coherent r l}
    \ket{\alpha_r} \equiv e^{\alpha_r a^\dagger_r-\alpha^*_ra_r} \ket{n_r\!=\!0}, \qquad
    \ket{\alpha_l} \equiv e^{\alpha_l a^\dagger_l-\alpha^*_la_l} \ket{n_l\!=\!0}.
\end{align}
These states transform under the exponential flow of $C_\perp$(for $q<0$) as

\begin{align}
    e^{-i C_\perp s}\ket{\alpha_r,\alpha_l} = e^{-i\frac{\mu_c^2}{2} s} \Ket{\alpha_r e^{-i\mu_c^2 s}, \alpha_l}
\end{align}
and they have mean value
\begin{align}
    \braket{C_\perp} = \mu_c^2\Bigl( |\alpha_r|^2 + \frac{1}{2} \Bigl) \simeq \mu_c^2|\alpha_r|^2, \label{MME}
\end{align}
and variance
\begin{align}
    \Delta C_\perp \equiv \sqrt{\Braket{(\Delta C_\perp)^2}} = \mu_c^2|\alpha_r|.
\end{align}
In the last equality in Eq.~\eqref{MME} we approximated for a coherent state with large mean occupation number $|\alpha_r|^2\gg1$. Note that, unlike for free Gaussians, one cannot pick the mean value and the variance of $C_\perp$ independently as they are ruled by the single parameter $|\alpha_r|$. Another relevant difference is that the (transverse) spatial spread of these coherent states is constant throughout evolution, whereas free particles undergo spatial diffusion.

Now, it is straightforward to solve the total Hamiltonian constraint \eqref{H  UMF} in terms of the complete set of commuting operators $\{H_C,P^0,P_z,N_r,N_l\}$ in $\mathcal{H}_{kin}$. The Hamiltonian $H$ acts on their diagonalized eigenstates as

\begin{align}
    H\ket{\epsilon,p^0,p_z,n_r,n_l} = \lambda\ket{\epsilon,p^0,p_z,n_r,n_l},
\end{align}
with
\begin{align}
    \lambda \equiv  \frac{\sqrt{U_\alpha U^\alpha}}{m+\epsilon} \Bigl[\! (m\!+\!\epsilon)^2 \!-\! (p^0)^2 \!+\! (p_z)^2 \!+\! \mu_c^2\Bigl(n_r \!+\! \frac{1}{2}\Bigl) \!\Bigl].
\end{align}
Thus, all the basis states spanning $\mathcal{H}_{phys}$ obey the dispersion relation

\begin{align}
    (m+\epsilon)^2 = (p^0)^2 - (p_z)^2 - 2m\omega_c\Bigl(n_r + \frac{1}{2}\Bigl),
\end{align}
which, for a positive mass spectrum, reduces to

\begin{align}
    (m+\epsilon) = \sqrt{(p^0)^2 - (p_z)^2 - 2m\omega_c\Bigl(n_r + \frac{1}{2}\Bigl)}.
\end{align}

By using this linearized dispersion relation in Eq.~ \eqref{free PSI}, we may now write a general history state as

\begin{widetext}
    
\begin{align}
    \ket{\Psi}\!\rangle = \int d\tau \mu_C \ket{\tau} \int d\mu(\epsilon) &\frac{d^2p_\parallel}{(2\pi)^2} \sum_{n_r,n_l} e^{i\epsilon\tau} \Psi(\epsilon,p^0,p_z,n_r,n_l)  \delta\bigl( m+\epsilon - \sqrt{(p^0)^2 - (p_z)^2 - \mu_c^2( n_r + 1/2)} \bigl) \ket{\epsilon,p^0,p_z,n_r,n_l}
\end{align}

\end{widetext}
Now, similarly to the case of a free clock, we want to choose a suitable initial state to evaluate our time evolution. We will again pick a clock with a continuum bounded spectrum $\sigma[H_C] = [-\Omega,\Omega]$, prepared in a state satisfying: 

$\bullet$ Gaussian in $z$-component of the momentum, centered at $p_z=0$ and with spread $\Delta_z$.

$\bullet$ Coherent state $\ket{\alpha_r, \alpha_l}$ for the transverse direction. However, since the left-polarized quanta do not contribute to the total energy, we simply take $\alpha_l=0$.

$\bullet$ Gaussian in $p^0$ with spread $\Delta_0$. However, rather than being centered at $p^0=m$, it will be centered at the average energy of our coherent state, $E=E(\alpha_r)$, given by
\begin{align}
    E(\alpha_r) = \sqrt{m^2 + \braket{C_\perp}} \simeq \sqrt{m^2 +\mu_c^2|\alpha_r|^2},
\end{align}
where the approximation in the above equation is for $|\alpha_r|\gg 1$. Thus, our initial state reads

\begin{widetext}
    
\begin{align}
    \ket{\psi(\tau\!=\!0)} =N \sum_{n_r=0}^\infty e^{-\frac{|\alpha_r|^2}{2}} \int \frac{dp^0 dp_z}{2\pi}\ket{p^0,p_z;n_r,n_l\!=\!0} \frac{(\alpha_r)^{n_r}}{n_r!}
    e^{-\frac{(p_0-E)^2}{\Delta_0^2}}e^{-\frac{p_z^2}{\Delta_z^2}} \, \chi\Bigl( \sqrt{p_\parallel^2-\mu_c^2(n_r+1/2)} - m \Bigl), \label{initial mag}
\end{align}
where $p_\parallel^2  \equiv (p^0)^2-p_z^2$ and $N$ is a normalization constant. From \eqref{initial mag} we can write its  time-evolution

\begin{align}
    \ket{\psi(\tau)}= N \sum_{n_r=0}^\infty e^{-\frac{|\alpha_r|^2}{2}} \int \frac{dp^0 dp_z}{2\pi}&\ket{p^0,p_z;n_r,n_l\!=\!0} \frac{(\alpha_r)^{n_r}}{n_r!}
    e^{-\frac{(p_0-E)^2}{\Delta_0^2}}e^{-\frac{p_z^2}{\Delta_z^2}} \nonumber \\
     &\times e^{-i\bigl[m- \sqrt{p_\parallel^2-\mu_c^2(n_r+1/2)} \bigl]\tau} \,\chi\Bigl( \sqrt{p_\parallel^2-\mu_c^2(n_r+1/2)} - m \Bigl). \label{UMF exact psi(tau)}
\end{align}

However, just as in the case of a free particle, it is illuminating to look at this expression for energies well within the bulk (far from the edges) of the spectrum of $H_C$ and in a weakly relativistic approximation. A sufficient condition for this regime is that $\Delta_0, \Delta_z, \mu_c|\alpha_r|^\frac{1}{2} \ll \Omega < m$, yielding

\begin{align}
    \ket{\psi(\tau)} =N \biggl( \int dt dz \ket{t,z} e^{iEt} e^{-\frac{\Delta_0'^2(\tau)}{2}\bigl[t - \bar{\gamma}\tau\bigl]^2} e^{-\frac{\Delta_z'^2(\tau)}{2}z^2} \biggl) 
    \otimes \ket{ \alpha_r e^{-i\omega_c\tau}, \alpha_l=0  }e^{i|\alpha_r|^2\omega_c\tau} \label{app NR UMF psi(tau)}
\end{align}
\end{widetext}
for the leading order approximation for our evolved state~ \eqref{UMF exact psi(tau)}. Here, we have Fourier-transformed the $p^0$ and $p^z$ variables to write the state in terms of the states $|t, z\rangle$. We also have defined $\bar{\gamma} \equiv E(\alpha_r)/m$, which we associate to an effective velocity $\bar{v}\equiv\sqrt{E^2(\alpha_r)-m^2}/E(\alpha_r)$, and a temporal spread

\begin{align}
    \Delta_0'^2(\tau) = \frac{\Delta_0^2}{1-i\Delta_0^2\,\bar{v}^2\bar{\gamma}^2\tau/m}.
\end{align}
In contrast to the case of an inertial particle, the temporal spread also diffuses with proper time, although this diffusion only appears as a higher-order relativistic correction: $\Delta'_0(\tau) = \Delta_0^2 \bigl( 1 + \mathcal{O}(\bar{v}^2) \bigl)$.


\end{document}